\documentclass[10pt,journal,twoside]{IEEEtran}

\usepackage{cite}
\usepackage{url}
\usepackage[colorlinks,linkcolor=blue,anchorcolor=green,citecolor=red, urlcolor=red, backref=false]{hyperref}

\usepackage{graphicx}
\usepackage{subfigure}
\usepackage{verbatim}
\usepackage{amsmath}
\usepackage{amsfonts}
\usepackage{amssymb}
\usepackage{dsfont}
\usepackage{setspace}
\usepackage{algorithm}

\usepackage{diagbox} % 加载tableI宏包
\usepackage{amssymb} % 加载tableI宏包
\usepackage{bm}

\usepackage{flushend}

\usepackage{multicol}

\usepackage{booktabs}

\usepackage{multirow}

\usepackage{threeparttable}
\usepackage[usenames,dvipsnames]{color}
\usepackage{colortbl}
\usepackage{caption}

\begin{document}

\title{De-Pois: An Attack-Agnostic Defense against\\Data Poisoning Attacks}
\author{Jian~Chen,~\IEEEmembership{Student Member,~IEEE},
~Xuxin~Zhang,
~Rui~Zhang,~\IEEEmembership{Member,~IEEE},\\
~Chen~Wang,~\IEEEmembership{Senior Member,~IEEE},
and~Ling~Liu,~\IEEEmembership{Fellow,~IEEE}
\IEEEcompsocitemizethanks{
\IEEEcompsocthanksitem This work was supported in part by the National Natural Science Foundation of China under Grants 61872416, 52031009, 62002104 and 62071192; by the Fundamental Research Funds for the Central Universities of China under Grant 2019kfyXJJS017; by the special fund for Wuhan Yellow Crane Talents (Excellent Young Scholar); and by the fund of Hubei Key Laboratory of Transportation Internet of Things under Grants 2019IOT004.
Ling Liu's research is partially support by the National Science Foundation under Grants NSF~2038029,
NSF~1564097, and an IBM faculty award. \emph{(Corresponding author: Chen Wang.)}
\IEEEcompsocthanksitem J. Chen, X. Zhang and C. Wang are with
the Internet Technology and Engineering R\&D Center (ITEC),
School of Electronic Information and Communications, Huazhong University of Science and Technology, Wuhan 430074, China. Email: \{jianchen, xuxinz, chenwang\}@hust.edu.cn.
\IEEEcompsocthanksitem R. Zhang is with Hubei Key Laboratory of Transportation Internet of Things, School of Computer Science and Technology, Wuhan University of Technology, Wuhan 430070, China. Email: zhangrui@whut.edu.cn.
\IEEEcompsocthanksitem L. Liu is with College of Computing, Georgia Institute of Technology, Atlanta, GA 30332-0765, USA. Email: ling.liu@cc.gatech.edu.
}
}

\markboth{IEEE Transactions on Information Forensics and Security}
{J. Chen \MakeLowercase{\textit{et al.}}: De-Pois: An Attack-Agnostic Defense against Data Poisoning Attacks}

\maketitle

\begin{abstract}
Machine learning techniques have been widely applied to various applications.
However, they are potentially vulnerable to data poisoning attacks, where sophisticated attackers can disrupt the learning procedure by injecting a fraction of malicious samples into the training dataset.
Existing defense techniques against poisoning attacks are largely {\em attack-specific}: they are designed for one specific type of attacks but do not work for other types, mainly due to the distinct principles they follow.
Yet few general defense strategies have been developed.
In this paper, we propose De-Pois, an {\em attack-agnostic} defense against poisoning attacks.
The key idea of De-Pois is to train a mimic model the purpose of which is to imitate the behavior of the target model trained by clean samples.
We take advantage of Generative Adversarial Networks (GANs) to facilitate informative training data augmentation as well as the mimic model construction.
By comparing the prediction differences between the mimic model and the target model, De-Pois is thus able to distinguish the poisoned samples from clean ones, without explicit knowledge of any ML algorithms or types of poisoning attacks.
We implement four types of poisoning attacks and evaluate De-Pois with five typical defense methods on different realistic datasets.
The results demonstrate that De-Pois is effective and efficient for detecting poisoned data against all the four types of poisoning attacks, with both the accuracy and F1-score over $\mathbf{0.9}$ on average.
\end{abstract}

\begin{IEEEkeywords}
Machine learning, data poisoning attack, attack-agnostic defense, generative adversarial network.
\end{IEEEkeywords}

\section{Introduction}\label{Introduction}
Machine learning (ML) has become a significant part of numerous systems and applications~\cite{RN208,7873244}. Despite the outstanding effectiveness of machine learning algorithms in many prediction and decision making tasks, recent studies show that ML algorithms are susceptible to potential security threats. For instance, attackers can steal the private information of ML models in model extraction attacks~\cite{tramer:2016stealing,shokri:2017membership}, obtain the private data of the training dataset in model inversion attacks~\cite{fredrikson:2015model,hitaj:2017deep}, induce misclassification in the testing time in evasion attacks~\cite{biggio:2013evasion,8949445}, or influence the training dataset to alter the prediction results of ML models in data poisoning attacks ~\cite{Battista-et-al:poisoning,alfeld:2016data,mei:2015using,munoz:2017towards,koh:2018stronger}.

\begin{figure*}[t]
	\centering
	\subfigure[Unpoisoned classifier.]
    {\label{fig:a}\includegraphics[width=45mm]{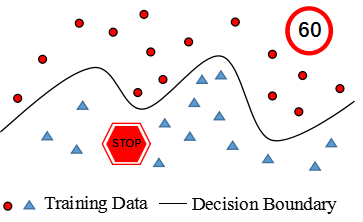}}\hspace{6mm}
	\subfigure[Poisoned classifier.]
    {\label{fig:b}\includegraphics[width=45mm]{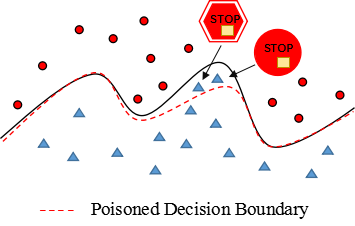}}\hspace{6mm}
	\subfigure[Poisoned classifier.]
    {\label{fig:c}\includegraphics[width=45mm]{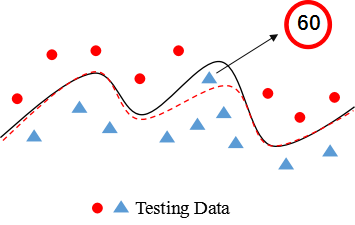}}
    \caption{Illustration of the data poisoning attack. (a) The classifier could classify the training dataset correctly. (b) Attackers add poisoned samples to the training dataset to manipulate the decision boundary. (c) Some testing data is misclassified in the testing phase due to the poisoning attack (the ``stop" sign is misjudged as the ``speed limit").}
	\label{fig:poison}
\end{figure*}

In this paper, we focus on the data poisoning attacks, where sophisticated attackers could disrupt the ML procedure by injecting a fraction of malicious samples into the training dataset.
Such vulnerabilities may pose serious risks to various security-critical domains such as self-driving cars~\cite{gu:2017badnets}, biometric identity recognition~\cite{chen-et-al:2017targeted} and computer vision~\cite{8854834}.
For instance, attackers may add stop signs with particular stickers into the training data to manipulate the decision boundary, so that the traffic sign classifier will misjudge the ``stop" as the ``speed limit" in the testing phase (c.f. Fig.~\ref{fig:poison}), which could potentially cause self-driving cars to maintain steering without stopping for obstacle avoidance.

To counter poisoning attacks, several defense techniques have been investigated recently ~\cite{paudice:2018detection,liu:2018fine,carnerero:2020regularisation}.
However, these defenses are largely {\em attack-specific}: they are designed for one specific type of attacks but may not work well for other types, mainly due to the distinct principles they follow. For example, Peri et al.~\cite{peri:2019deep} mitigate targeted clean-label attacks (a type of poisoning attacks) by identifying poison samples from its $k$ nearest neighbors in the feature space. It could detect correctly-labeled and minimally-perturbed samples but fails when poisoned data is generated by gradient-based method~\cite{Yang:2017GenerativePA} where the labels of samples are elaborately manipulated. In another case, Jagielski et al.~\cite{Jagielski:2018ManipulatingML} propose to defend against poisoning attacks in the context of regression; unfortunately, this approach cannot be applied to attacks in the context of classification~\cite{Biggio:2011SupportVM}.
To date, few general defense strategies have been developed against such poisoning attacks.

To fill this gap, in this paper, we propose De-Pois, an {\em attack-agnostic} approach for defense against poisoning attacks. De-Pois is motivated by a fundamental practice in the way attackers launch poisoning attacks: poisoned samples are injected to manipulate the decision boundary of the target model trained by clean samples, and there exists a difference between the feature spaces of poisoned samples and clean samples. Therefore, the key idea of De-Pois is to train a mimic model whose purpose is to imitate the behavior of the target model. Using the constructed mimic model, it is then straightforward to distinguish the poisoned samples from the clean ones by comparing the prediction differences between the mimic model and the target model.

Though the basic idea is simple, there are two major challenges to be addressed.
The first challenge is how to obtain sufficient valid training data for the mimic model, which is expected to have a similar distribution of the clean data.
In many practical scenarios, especially in user-provided data systems, it is feasible to obtain only a small number of clean data from trusted data sources (e.g., creditworthy users)~\cite{miao:2018attack}, which is far from being enough.
The second challenge is how to train an effective mimic model, which can achieve comparable prediction performance with the target model, since the structure or the hyper parameters of the target model are unknown in advance.
Without the effective mimic model, it is infeasible to obtain precise prediction difference to recognize the poisoned samples from the clean ones.

To tackle the above challenges, De-Pois incorporates two novel designs, namely, synthetic data generation and mimic model construction, based on Generative Adversarial Networks (GAN)~\cite{Goodfellow:2014}.
Specifically, De-Pois modifies conditional GAN (cGAN)~\cite{Mirza-Osindero:2014CGAN} to better understand the underlying distribution of the clean data, so that sufficient valid training data can be generated from a small fraction of trusted clean data.
%, and devises an authenticator to supervise the data augmentation process, .
A conditional version of Wasserstein GAN gradient penalty (WGAN-GP)~\cite{gulrajani:2017improved} is further adopted to learn the distribution of predictions of the augmented training data, yielding a mimic model with similar prediction functionality with the target model.
In this way, De-Pois can finally employ the mimic model to recognize the poisoned data from testing samples by comparing the difference between the mimic model's output and a properly determined detection boundary.

The contributions of this paper are summarized as follows.
\begin{itemize}
  \item To the best of our knowledge, De-Pois is the first generic method for defending against poisoning attacks. De-Pois provides its protection without explicit knowledge of any ML algorithms or types of poisoning attacks, and can be deployed for protecting both classification and regression tasks.
  \item We take advantage of cGANs to map latent space representation to the distribution of the clean data, which facilitates informative training data augmentation.
      We further employ conditional WGAN-GP to fit the Wasserstein distance between augmented data and clean data for the mimic model construction.
  \item We evaluate the effectiveness of De-Pois against four types of poisoning attacks, and compare De-Pois with five typical defense methods on four realistic datasets for different ML tasks. The results demonstrate that De-Pois is very effective and efficient for detecting poisoned data against all the four types of poisoning attacks, with both the accuracy and F1-score over $0.9$ on average.
\end{itemize}

%RANSAC~\cite{Fischler:1987RandomSC}  &
\begin{table*}[t]
	\centering
	\begin{threeparttable}
		{\fontsize{9.5}{12}\selectfont
			\caption{Summary of typical defenses against four types of poisoning attacks.}  \label{tab:feasibility}
			\setlength{\tabcolsep}{1.7mm}{	
				\begin{tabular}{llllllll}		
					\hline
					\diagbox[width=10em,trim=l]{Attack}{Defense} &  Deep-$k$NN~\cite{peri:2019deep}  & CD~\cite{Steinhardt:2017CertifiedDF}  &DUTI~\cite{zhang:2018training} & TRIM~\cite{Jagielski:2018ManipulatingML} &  Sever~\cite{diakonikolas:2019sever} & \textbf{De-Pois} \\
					\hline
				TCL-attack~\cite{zhu:2019transferable} & \multicolumn{1}{c}{$\surd$}& \multicolumn{1}{c}{$\bigcirc$} & \multicolumn{1}{c}{$\bigcirc$}                  & \multicolumn{1}{c}{$\bigcirc$}     &   \multicolumn{1}{c}{$\bigcirc$}     &\multicolumn{1}{c}{$\surd$}        \\
					\hline			
					pGAN-attack~\cite{munoz:2019poisoning}& \multicolumn{1}{c}{$\bigcirc$}&\multicolumn{1}{c}{$\surd$}                    &  \multicolumn{1}{c}{$\bigcirc$}                & \multicolumn{1}{c}{$\bigcirc$}     &  \multicolumn{1}{c}{$\bigcirc$}      &\multicolumn{1}{c}{$\surd$}         \\
					\hline
					LF-attack~\cite{Biggio:2011SupportVM}& \multicolumn{1}{c}{$\bigcirc$}  & \multicolumn{1}{c}{$\surd$}                    & \multicolumn{1}{c}{$\surd$}              & \multicolumn{1}{c}{$\bigcirc$}        &\multicolumn{1}{c}{$\surd$}    &\multicolumn{1}{c}{$\surd$}         \\
					\hline
					R-attack~\cite{Jagielski:2018ManipulatingML}& \multicolumn{1}{c}{$\bigcirc$}  &  \multicolumn{1}{c}{$\bigcirc$}                     &  \multicolumn{1}{c}{$\surd$}             & \multicolumn{1}{c}{$\surd$}       &\multicolumn{1}{c}{$\surd$}    &\multicolumn{1}{c}{$\surd$}        \\
					\hline
				\end{tabular}
		}}
		%\vspace{-2pt}
		\begin{tablenotes}
			\item[]{\small Effective defense: $\surd$ \qquad Ineffective defense: $\bigcirc$}
		\end{tablenotes}
		
	\end{threeparttable}
	
\end{table*}

The remainder of this paper is organized as follows.
Section~\ref{Related Work} briefly introduces existing poisoning attacks and corresponding defenses.
Section~\ref{Threat Model} describes the threat model and defense capability.
Section~\ref{Proposed De-Pois} provides design details of De-Pois, followed by some discussions in Section~\ref{Discussion}.
Section~\ref{Experimental evaluation} presents experiment results, and finally Section~\ref{conclusion} concludes this paper.
The code of De-Pois has been released for reproducibility purposes\footnote{\url{https://www.dropbox.com/s/rkwpjd8chci0b3f/De-Pois-Code.zip?dl=0}}.

\section{Related Work} \label{Related Work}
In this section, we briefly review four types of poisoning attacks and their typical defense techniques.

\subsection {Targeted Clean-Label Poisoning Attack (TCL-attack)} \label{TCL-attack}
Targeted clean-label attack is a type of poisoning attack in which the attacker adds cleanly-labeled, minimally-perturbed data into the training samples, causing ML model to misclassify a specific test sample at testing time. For example, Shafahi et al.~\cite{shafahi:2018poison} craft clean-label poisoning data via feature collisions. In particular, they craft the poison instance that is close to the target sample in feature space while staying close to the base instance at the same time. More recently, a stronger targeted clean-label method is proposed~\cite{zhu:2019transferable}, which looks for a looser constraint on the poisoned samples to avoid obvious patterns of the target in the attack, and thus can prevent the poisoned samples from being easily detected.

To defend against such attacks, Deep-$k$NN~\cite{peri:2019deep} removes malicious samples by comparing the class labels of each testing sample with its $k$ neighbors, based on the intuition that poisoned samples have different feature representations than those of clean samples. In this sense, a sample would be regarded as poisoned if the majority of the $k$ samples surrounded by the testing sample do not share the same class label as itself.

\subsection {Poisoning Attack with GAN (pGAN-attack)} \label{pGAN-attack}
Mu{\~n}oz-Gonz{\'a}lez et al.~\cite{munoz:2019poisoning} introduce a GAN-based poisoning attack called pGAN-attack to craft adversarial training examples. This attack consists of the generator, the discriminator and the classifier, and utilizes the discriminator to distinguish the pristine and the generated poisoning data. A hyperparameter $\alpha$ is used to trade off the detectability and effectiveness of the poisoned samples, where a higher value of $\alpha$ indicates more chances for the crafted samples to evade the detection.

As an effective defense method against pGAN-attack, certified defense (CD)~\cite{Steinhardt:2017CertifiedDF} first filters outliers outside the estimated feasible set and then minimizes a margin-based loss for the rest of the data. Afterwards, CD explores certified defenses by computing data-dependent upper bounds on the testing loss, where samples poisoned by pGAN-attack with longer distance from the true class centroids can be identified.

\subsection {Label-Flipping Attack (LF-attack)} \label{Adversarial Label Attack}
Biggio et al.~\cite{Biggio:2011SupportVM} develop an adversarial label flips method to attack SVM. Particularly, they increase the probability of flipping the label of input data which are categorized with high confidence, and in this way the attacking success rate is improved.

Label sanitization (LS)~\cite{Paudice:2018LabelSA} is then proposed to mitigate label-flipping attacks, which relies on the observation that poisoned samples are far away from the decision boundary of SVM and thus have more chance to be relabelled.

Meanwhile, Zhang et al.~\cite{zhang:2018training} devise a stronger defense, dubbed DUTI, against the label-flipping attack. Specifically, DUTI formulates a bi-level optimization problem to handle the teaching task. Given a small portion of the trusted items, DUTI would learn to find the potentially corrupted labels, and then gives them to a domain expert for further inspection to identify outliers.

\subsection {Regression Attack (R-attack)} \label{Regression Attack}
Jagielski et al.~\cite{Jagielski:2018ManipulatingML} mostly focus on poisoning attacks for linear regression. They formulate the poisoning attack as a bi-level optimization problem and Karush-Kuhn-Tucker conditions are employed to solve the non-convex problem. Furthermore, a statistical-based poisoning attack is put forward for black-box attacks in which attackers could query the target model to find statistical features of the training samples.

To counter such attacks, TRIM~\cite{Jagielski:2018ManipulatingML} trains a regression model on a subset of samples with poisoned data and iteratively estimates the residuals. It is indicated that the subset of samples with the smallest residual can be identified as pristine.
Also, DUTI~\cite{zhang:2018training} can be applied to solving poisoning attack under regression setting.

\vspace{5pt}
\textbf{Summary.}
Existing defense techniques against poisoning attacks are {\em attack-specific} and can only defend some specific type of poisoning attacks (c.f. Table~\ref{tab:feasibility}).
In particular, these defenses identify poisoned data following distinct analysis of either specific training loss which contains class label or regression value, or using nearest neighbors techniques.
Thus, it is difficult for one type of defense to detect poisoned data when it comes to attacks for different learning tasks.
It is noticed that a recent model-agnostic defensive mechanism dubbed Sever~\cite{diakonikolas:2019sever} can be adopted to defend different types of attacks, but may fail for some deep neural network (DNN) targeted attacks such as TCL-attack and GP-attack, due to the difficulty in the model fitting of diverse unknown DNN models.
In practice where the defender can hardly get aware in advance of which type of attacks has been carried out, those attack-specific defenses may become ineffective any more.
In contrast, De-Pois is a {\em generic} and {\em attack-agnostic} defense approach which can work effectively for all the aforementioned types of attacks, and is more promising in real-world scenarios.

\section {Threat Model and Defense Capability} \label{Threat Model}
In this section, we first provide a detailed threat model for the poisoning attacks described in the previous section and then describe the power of defenders.
The threat model consists of the attacker's goal, the attacker's knowledge of the target model, as well as the attacker's capability of how to influence the training data.
The defender's capability, on the other hand, is largely related to the obtained knowledge under the threat model.

\textbf{Attacker's goal.}
The attacker's goal can be categorized into two classes~\cite{biggio:2018wild}: the poisoning availability attack, which aims to affect the model's prediction performance indiscriminately, e.g., pGAN-attack~\cite{munoz:2019poisoning}, LF-attack~\cite{Biggio:2011SupportVM} and R-attack~\cite{Jagielski:2018ManipulatingML},
and the poisoning integrity attack, where the attacker aims to bring about specific prediction errors for some desired testing samples, e.g., TCL-attack~\cite{zhu:2019transferable}.

\textbf{Attacker's knowledge.}
The attacker may have different levels of knowledge of the target ML model. In white-box attacks~\cite{munoz:2019poisoning,Biggio:2011SupportVM,Jagielski:2018ManipulatingML}, the attacker is expected to know the unadulterated training data, the feature values of each sample, the ML algorithm, and the parameters of the trained model. In the black-box attack~\cite{zhu:2019transferable}, the attacker has no knowledge of the aforementioned information but can acquire a substitute dataset with a similar distribution of the original training data.

\textbf{Attacker's capability.}
The attacker's capability is typically refers to how and to what extend the attacker controls the training data. Normally, the attacker can modify the feature values and labels of the training data~\cite{Jagielski:2018ManipulatingML}, but one may also modify only feature values~\cite{munoz:2019poisoning,zhu:2019transferable} or only labels~\cite{Biggio:2011SupportVM}). On the other hand, the attacker is often constrained by upper bounding the number of poisoning samples, and the ratio of poisoning samples below $30\%$ is often mandatory.

\textbf{Defender's Capability.}
The defender could obtain different levels of knowledge of the target model and the training data, according to different threat models.
For example, Deep-$k$NN~\cite{peri:2019deep} assumes the availability of the ground-truth labels to compare the class labels with each sample's $k$ neighbors.
CD~\cite{Steinhardt:2017CertifiedDF} supposes the outliers do not have a strong effect on the target model in order to make an approximation about the upper bounds on the testing loss across poisoning attacks under non-convex settings. TRIM~\cite{Jagielski:2018ManipulatingML} makes a strong assumption on the pre-knowledge of the ratio of poisoning samples which is controlled by the attacker, while Sever~\cite{diakonikolas:2019sever} makes underlying assumptions on small singular values of the gradients of the training data. Similar to DUTI~\cite{zhang:2018training}, De-Pois assumes to have access to partial trusted training data.
From the defender's perspective, one may need the access to corresponding resources so as to defend against different types of attacks in practice, and the defender's capability thus may change under different threat models.

\section{De-Pois Design} \label{Proposed De-Pois}
\begin{figure*}[t]
\centering
\includegraphics[scale=0.49]{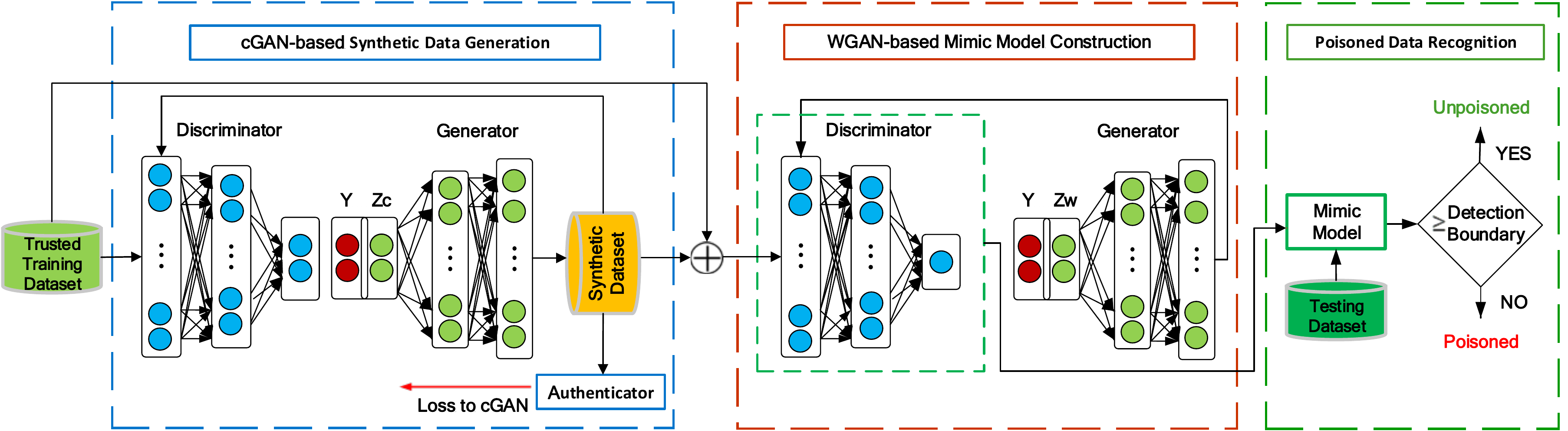}
\caption{
The framework of De-Pois, which contains three parts: the cGAN-based synthetic data generation, the WGAN-based mimic model construction and the poisoned data recognition.}
\label{fig:de_pois}
%\vspace{0.1cm}
\end{figure*}

We formulate our problem as follows:
for a training dataset $S_o = S_p \cup S_c$, which contains a potentially poisoned dataset $S_p$ and a clean dataset $S_c$, De-Pois aims to determine whether a sample $s\in S_o$ is in $S_p$ or not, given a small amount of (trusted) clean data $S_t\in S_o$.
De-Pois relies on the observation that poisoned samples are more likely to have different predictions than clean samples do.
Therefore, De-Pois tests out the poisoned samples by estimating their prediction difference, making use of a mimic model with a similar prediction behavior of the target model trained by $S_c$.
To this end, De-Pois mainly consists of three steps (c.f. Fig.~\ref{fig:de_pois}).

(1) {\bf Synthetic Data Generation.}
The first step of De-Pois is to generate sufficient synthetic training data with a similar distribution of $S_c$, in condition that only $S_t$ can be obtained in practice.
To better understand the underlying distribution of $S_c$, De-Pois leverages cGAN for data generation and devises an authenticator to supervise the data augmentation.

(2) {\bf Mimic Model Construction.}
After obtaining sufficient valid data, De-Pois next builds the mimic model by developing the conditional version of WGAN-GP to learn the distribution of predictions of the augmented training data.
When the training of the conditional WGAN-GP is completed, we regard its \emph{Discriminator} as our mimic model.

(3) {\bf Poisoned Data Recognition.}
Given the mimic model, De-Pois can thus employ a detection boundary to set apart the poisoned samples from clean ones. If the mimic model's output is lower than our detection boundary, the sample is then regarded as being poisoned.

\subsection {Synthetic Data Generation} \label{Synthetic Data Generation}

In general, our synthetic data generation module consists of two parts: a cGAN-based generator and an authenticator.

%\subsubsection{GAN} \label{GAN}
\subsubsection{cGAN-based Generator} \label{CGAN}
The original GAN contains two neural networks~\cite{Goodfellow:2014}: the Generator ($G$) and the Discriminator ($D$).
$G$ learns to generate synthesized samples $G(\boldsymbol z)$ which captures the distribution of training data $P_{data}(\boldsymbol x)$ from a prior noise distribution $\boldsymbol z$, while $D$ learns to distinguish real data samples $\boldsymbol x$ from $G(\boldsymbol z)$.
$G$ and $D$ are learned simultaneously to achieve the following min-max objective:
\begin{equation}
\begin{split}
\min_{G}\max_{D}V (G, D)=\mathbb{E}_{\boldsymbol x\sim P_{data}(\boldsymbol x)}[\log D(\boldsymbol x)]\\
+\,\mathbb{E}_{\boldsymbol z\sim P_{\boldsymbol z}(\boldsymbol z)}[\log(1-D(G(\boldsymbol z)))].
\end{split}
\end{equation}

It is noticed that in the original GAN, no control on modes of the generated data is employed. Therefore, it is not effective for GAN to guide the data generation process in this unconditioned generative model.
By making $G$ conditioned on additional information (e.g., the class labels), cGAN synthesizes new training samples by applying a random noise vector $\boldsymbol z_c$ and a condition constraint like class labels or other modalities to train the model~\cite{Mirza-Osindero:2014CGAN}.

To take advantage of cGAN, we thus propose to feed the additional information $\boldsymbol y$ into both the Generator and the Discriminator (i.e., $G_c$ and $D_c$) in cGAN, and generate samples conditioned on $\boldsymbol y$ in a supervised way. It is important to note that the regression value $\boldsymbol y$ adding an additive unimodal noise $\boldsymbol z_r$ (e.g., Gaussian noise) is conditioned for regression tasks in order to cover the whole set of outputs as much as possible.
The objective function can thus be as:
\begin{equation}
\begin{split}
L_{\textit{cGAN}}&=V (G_c, D_c)\\
&=\mathbb{E}_{\boldsymbol x\sim P_{data}(\boldsymbol x)}[\log D_c(\boldsymbol x|\boldsymbol y)]\\
&+\mathbb{E}_{\boldsymbol z_c\sim P_{\boldsymbol z_c}(\boldsymbol z_c)}[\log(1-D_c(G_c(\boldsymbol z_c|\boldsymbol y)))].
\end{split}
\end{equation}

\subsubsection {Authenticator} \label{Authenticator}
 Intuitively, cGAN can generate sufficient data. However, the generated data generally has less diversified expressions mainly due to the single distribution (e.g. Gaussian distribution) fed to $G_c$ and it is uncertain if cGAN can generate data with high fidelity and variety in low-volume data scenario~\cite{tran:2017bayesian}.

To generate more valid training data, we thus introduce an authenticator to supervise the data augmentation process in cGAN. In particular, we treat these newly synthesized samples generated by $G_{c}$ at each iteration as instances of missing latent variables which lie in the space of training data. We then calculate the loss $L_{A}$ between the authenticator's prediction output and the true class label or regression value for each synthesized sample. At last, we back-propagate the loss $L_{A}$ to the cGAN part, with the loss of $D_c$ (resp. $G_c$) as $L_{\textit{cGAN}}+L_{A}$ (resp. $L_{\textit{cGAN}}-L_{A}$).
It is noted that the idea of adopting an authenticator is motivated by the Bayesian DA model~\cite{tran:2017bayesian}, which deals with only classification tasks. We extend the authenticator to work with both classification and regression tasks, and improve the granularity of the training process.

Due to different poisoning attack tasks (i.e., classification and regression tasks), the authenticator calculates the loss $L_{A}$ in a different way.
For classification, the authenticator is designed as a convolutional neural network (CNN). We first get the output $\hat{\boldsymbol y}$ of the authenticator, and then use the cross entropy error function to calculate the loss for classification, which can be formulated by:
\begin{equation} \label{authen_c}
L_{A}=-\frac{1}{M_s} \sum_{i=1}^{M_s}\sum_{j=1}^{N_c}y_{i}^j\log(\hat{y_{i}^j}),
\end{equation}
where $\hat{y_{i}^j}$ is the prediction probability of the $i$th sample belonging to class $j$,
$N_c$ is the number of classes and $M_s$ is the number of synthesized sample at each iteration.
$y_{i}^j=1$ if the label of the $i$th sample belongs to class $j$; otherwise $y_{i}^j=0$.

For regression, the authenticator calls for a specific regression module (e.g., LASSO), and the loss is formulated by Equ.~(\ref{authen_r}) using mean square error (MSE) of each synthesized sample:
\begin{equation} \label{authen_r}
\displaystyle L_{A}=\frac{1}{M_s} \sum_{i=1}^{M_s}(y_{i}-\hat{y_{i}})^2,
\end{equation}
where $y_{i}$ represents the regression value of the $i$th sample and $\hat{y_{i}}$ is the prediction value of the authenticator for the $i$th sample at each iteration.

In this way, the authenticator encourages better distinction between the real data and the generated data, and thereby can enhance the data augmentation process.

\subsubsection {Synthetic Data Generation} \label{Synthetic Data Generation}

The cGAN training process of synthetic data generation involves two parts: for the discriminative part, we use the trusted clean samples $S_t$ as its input, aiming to minimize $L_{\textit{cGAN}}+L_{A}$. For the generative part, a noise prior $\boldsymbol z_c$ and additional information $\boldsymbol y$ are combined as its input, aiming to minimize $L_{\textit{cGAN}}-L_{A}$. In the synthetic data generation process, these two parts are optimized in an adversarial way as normally done in GAN.

In order to determine the parameters in cGAN-based synthetic data generation process, we employ Monte Carlo Expectation Maximization (MCEM) and run it iteratively, where we first utilize Monte Carlo (MC) to estimate the value of model's parameters based on the estimation results of the previous iteration and then update these parameters with stochastic gradient decent (SGD) at each iteration.

In the expectation-maximization (EM) algorithm, we estimate the parameters $\theta$ of our synthetic data generation model using the trusted clean data $S_t=\left\{ \boldsymbol s_{i} \right\}_{i=1}^{N_t}$ which corresponds to a given data $\boldsymbol s=(\boldsymbol x,\boldsymbol y)$, with $\boldsymbol x$ representing data sample, $\boldsymbol y$ denoting the true label (or regression value), and $N_t=|S_t|$. The training process can be formalized as the following optimization problem:
\begin{equation}
\theta^*=\arg\max_{\theta}\log p(\theta|\boldsymbol s).
\end{equation}

It is noticed that we cannot calculate the posterior $p(\theta|\boldsymbol s)$ directly in the absence of information such sa the prior and likelihood function.
So we increase the training data using synthesised data which can be represented by a latent variable $\boldsymbol z_{s}=(\boldsymbol x_{s},\boldsymbol y_{s})$, where $\boldsymbol x_{s}$ represents a synthesized data, and $\boldsymbol y_{s}$ refers to the associated class label (or regression value). Given both $\boldsymbol s$ and $\boldsymbol z_{s}$, we can estimate the augmented posterior $p(\theta|\boldsymbol s,\boldsymbol z_{s})$ at the $i$th iteration in E-step:
%To obtain the optimal parameters $\theta^*$, we first denote our data augmentation model parameters at the $i$th iteration by $\theta^i$.
\begin{equation}
\begin{split}Q(\theta,\theta^i)&=\mathbb{E}_{p(\boldsymbol z_{s}|\theta^i,\boldsymbol s)} [\log p(\theta|\boldsymbol s,\boldsymbol z_{s})].
\end{split}
\end{equation}
Then, M-step maximizes the $Q$ function in the next iteration:
\begin{equation}
\theta^{i+1}=\arg\max_{\theta}\, Q(\theta,\theta^i).
\end{equation}
We stop the training process once $||\theta^{i+1}-\theta^i||$ is sufficiently small, and obtain the optimal $\theta^*$ from the last iteration.
%However, it is difficult to compute the complicated integration in E-step.

We further employ MC strategy which uses large repeated random sampling to approximate the integration in E-step~\cite{Tanner:2012tool}. Furthermore, in M-step, we update $\theta^{i+1}$ by running SGD which uses only a sub-set of trusted clean data and augmented data in each iteration and can finally acquire the expected synthetic data generation model, leading to sufficient synthetic training data $S_s$ with a similar distribution of $S_c$.

Note that our aim is to obtain the augmented data $S_{\textit{aug}}$, such that $|S_{\textit{aug}}|=|S_t|+|S_s'|$ is comparable to $|S_o|$.
In most cases in our experiments, $|S_s|>|S_{\textit{aug}}|$, thus we randomly choose a subset of $S_s$ as $S_s'$.
In some cases when $|S_t|+|S_s|<|S_{\textit{aug}}|$, we can continue the training process to generate more data until $|S_s'|$ satisfies our setting.

\subsection {Mimic Model Construction} \label{Mimic Model Construction}

After obtaining $S_{\textit{aug}}$, De-Pois next aims to construct the mimic model which has similar prediction performance with the target model.
The idea is straightforward: we can regard the mimic model as functionally equivalent with the target model if the prediction outputs of the mimic model trained on $S_{\textit{aug}}$ are indistinguishable from those of the target one.

Simply using GAN to construct our mimic model seems feasible. However, we find that the original GAN suffers from training instability, mainly due to the discontinuity of the generator's parameters when minimizing the Jensen-Shannon (JS) divergence in practice.
To circumvent tractability issues, we propose a conditional version of WGAN-GP by making both the Generator and Discriminator parts of WGAN-GP conditioned on additional information $\boldsymbol y$.
In this manner, we could construct our mimic model better in a supervised way.

In the paradigm of WGAN-GP~\cite{gulrajani:2017improved}, the training instability issue is solved by penalizing on the norm of weights for random samples $\hat{\boldsymbol x}\sim P_{\hat{\boldsymbol x}}$ from its Discriminator network to make it satisfy Lipschitz constraint, and the gradient penalty is directly added to the Wasserstein distance~\cite{arjovsky:2017wasserstein}. The objective function is formulated as:
\begin{equation} \begin{split} L_{\textit{WGAN-GP}}=\mathbb{E}_{\widetilde{\boldsymbol x}\sim P_{g}}[D_{w}(\widetilde{\boldsymbol x})]-\mathbb{E}_{\boldsymbol x\sim P_{r}}[D_{w}(\boldsymbol x)]\\
+\lambda \mathbb{E}_{\hat{\boldsymbol x}\sim P_{\hat{\boldsymbol x}}}[(||\nabla_{\hat{\boldsymbol x}}D_{w}(\hat{\boldsymbol x})||_{2}-1)^2],\end{split}\end{equation}	
where the last item represents the penalty on the gradient norm. $P_{r}$ and $P_{g}$ represent the real and generated data distribution, respectively. $P_{\hat{\boldsymbol x}}$ denotes a uniform sampling distribution sampled from $P_{r}$ and $P_{g}$.

The WGAN-GP model provides us a more stable training environment. However, as mentioned before, it also suffers from inefficiency in data generation process under unconditioned generative model. In addition, the deep learning model will lead to training overfitting in low-data scenario. By introducing additional information into the WGAN-GP setting, we feed $\boldsymbol y$ into both $G_w$ and $D_w$ and are thus able to mimic the target model better in a supervised way.
Accordingly, the objective function of our mimic model combining both WGAN-GP and cGAN can be modified as:
\begin{equation} \begin{split} L_{\textit{cWGAN-GP}}=\mathbb{E}_{\widetilde{\boldsymbol x}\sim P_{g}}[D_{w}(\widetilde{\boldsymbol x}|\boldsymbol y)]-\mathbb{E}_{\boldsymbol x\sim P_{r}}[D_{w}(\boldsymbol x|\boldsymbol y)]\\
+\lambda \mathbb{E}_{\hat{\boldsymbol x}\sim P_{\hat{\boldsymbol x}}}[(||\nabla_{\hat{\boldsymbol x}}D_{w}(\hat{\boldsymbol x}|\boldsymbol y)||_{2}-1)^2].\end{split}\end{equation}

During the training process of our mimic model, we alternately optimize $D_w$ and $G_w$. After adequate epochs of training, and when the objective of both parts are converged, we complete the mimic model construction and regard $D_{w}$ as our expected mimic model.

\subsection {Poisoned Data Recognition} \label{Poisoned Data Recognition}
Using the mimic model, De-Pois can thus find out the poisoned samples in a straightforward way:
simply setting a detection boundary, and comparing the value between the mimic model's output and the detection boundary.
If the output's value is lower than our detection boundary, the sample is then regarded as being poisoned. Otherwise, the sample belongs to unpoisoned.

To be more concrete, in the context of both classification and regression tasks, each sample $x$ is fed into the mimic model and its prediction value $y_\textit{pre}$ is output. In our mimic model, $y_\textit{pre}$ represents the Wasserstein distance between the generated sample and the real sample.
In general, the prediction value $y_\textit{pre}$ of a clean sample is larger than that of a poisoned one. Thus the poisoned samples can be recognized from the clean ones given a proper detection boundary.
Since in practice we cannot know the clean samples $S_c$ in advance, and the distribution of our augmented samples $S_{\textit{aug}}$ is designed to be similar to that of $S_c$, we thus utilize $S_{\textit{aug}}$ to determine the detection boundary.

It is observed that the distribution of the prediction values of $S_{\textit{aug}}$ passing $D_w$ (denoted as $P_{\scriptscriptstyle S_{\textit{aug}}}$) almost fits a normal distribution. Thus it is feasible to determine the detection boundary only using $S_{\textit{aug}}$ without the knowledge of the poisoned samples. So in order to properly determine the detection boundary, we first obtain the mean $\mu$ and standard deviation $\sigma$ of $P_{\scriptscriptstyle S_{\textit{aug}}}$. After that, we calculate $z$-scores by standardizing the
distribution of the prediction values of testing samples passing $D_w$ (denoted as $P_{\scriptscriptstyle S_{\textit{test}}}$), which can be formulated by $\tilde{z}^i=\frac{y_{pre}^i-\mu}{\sigma}$,
where $y_{pre}^i$ and $\tilde{z}^i$ are the prediction value and the corresponding $z$-score of the $i$th testing sample.
We can then acquire $N$ $z$-scores corresponding to $N$ testing samples, which enables us to distinguish the difference in $P_{\scriptscriptstyle S_{\textit{test}}}$ with different means and standard deviations.
%Also we assume samples to be poisoned if the corresponding $z$-score is too far from zero and lower than a specific value $z_s$. Besides, the standard normal distribution table helps us to determine $z_s$.
From practical rule, we regard testing samples with unilateral confidence interval higher than $z_s$ as clean ones given the value of level of significance (e.g., $0.05$ in our experiments). Then we can obtain corresponding proper $z_s$ by looking up the standard normal distribution table (e.g., $z_s=-1.96$ corresponding to the level of significance $0.05$).

Finally, we can establish the detection boundary $y_\textit{thr}=z_s\times\sigma+\mu$ and compare it to a sample's prediction $y_\textit{pre}$. If the following inequation holds,
\begin{equation}
y_\textit{pre}<y_\textit{thr},
\end{equation}	
we then regard this sample as being poisoned.

After testing all the samples in $S_o\setminus S_t$, we can identify the poisoned data, which is further excluded from $S_o$.
In this way, De-Pois can be utilized as a filter prior to training the ML model so that the trained model will not be affected by the poisoned samples.

\section{Discussion} \label{Discussion}
In this section, we discuss some issues regarding the applicability of De-Pois from the following two aspects: application scope, as well as defending scope.
	
\subsection{Application Scope of De-Pois} \label{Applicability of De-Pois}

There are generally two strategies for defending against poisoning attacks. The first one is to devise a robust optimization algorithm which focuses on reducing the impact of the poisoning samples on the prediction performance of the target model. For example, in TRIM~\cite{Jagielski:2018ManipulatingML} the model is trained on the dataset with the lowest residuals and the impact of poisoning samples is weakened iteratively. Similar strategy is also adopted by Sever~\cite{diakonikolas:2019sever} and DUTI~\cite{zhang:2018training}.
The other strategy is to perform poisoned data removal on the training data and then employ proper ML algorithms on the sanitized data. Under this situation, the defense methods need to find out the poisoned data precisely, e.g., CD~\cite{Steinhardt:2017CertifiedDF} and our De-Pois.
In some scenarios, identifying the poisoned data is necessary and sometimes crucial, especially for tracing the poisoning sources.
For example, in intelligent crowd-sensing systems that rely on continuously collecting samples from the physical world, it is more desirable to find out and remove the malicious participators.
In this sense, De-Pois can be used not only to defend poisoning attacks, but also to facilitate system ML model developers fixing systematic training set errors caused by data poisoners.

\subsection{Defending Scope of De-Pois} \label{Attack Applicability}
In Section~\ref{Related Work}, we have described several types of poisoning attacks which De-Pois is able to defend against (c.f. Table~\ref{tab:feasibility}).
Now that De-Pois is a GAN-based approach, one question is naturally raised: will De-Pois still work if an attacker adopts a similar GAN-based approach to generate the poisoning samples?
Here, we introduce another interesting GAN-based method called GP-attack~\cite{Yang:2017GenerativePA} that illustrates how DNN could be used to generate malicious training samples. A generative method was proposed to accelerate the poisoned sample generation, using an autoencoder to generate poisoned samples and a discriminator to calculate the loss with GAN.
For the defense against GP-attack, CD~\cite{Steinhardt:2017CertifiedDF} can be an option as it can remove poisoned data far away from the true class centroids. Our De-Pois also works and is validated by experiments to be more effective than CD (c.f.~Section~\ref{GP-attack_experiment}).

\section{Performance Evaluation} \label{Experimental evaluation}
\subsection{Experiment Setup} \label{Experimental Setup}

\subsubsection{Datasets} \label{Dataset Description}
We perform experiments using four datasets in different domains including hand-written digit
	recognition, image classification, non-linear binary classification, and house sale price prediction, which are also adopted in existing poisoning attacks and defenses~\cite{peri:2019deep,gao:2019strip,Udeshi:2019ModelAD,Biggio:2011SupportVM,Jagielski:2018ManipulatingML}.

\textbf{MNIST}\footnote{http://yann.lecun.com/exdb/mnist}. This dataset consists of $28\times28$ gray-scaled images of the handwritten digits from $0$ to $9$, along with a training dataset of $60,000$ images and a testing dataset of $10,000$ images.

\textbf{CIFAR-10}\footnote{https://www.cs.toronto.edu/~kriz/cifar.html}. This dataset contains $60,000$ $32\times32$ color images in $10$ different classes, with $50,000$ training images and $10,000$ testing images. These $10$ different classes include cats, deer, dogs, frogs, airplanes, cars, horses, ships, birds, and trucks.

\textbf{Fourclass Dataset}\footnote{https://www.csie.ntu.edu.tw/\%7ecjlin/libsvmtools/datasets/binary.html}. This dataset has preprocessed into two-class stored in LIB-SVM repository and is not linearly separable. It contains $862$ samples in a two-dimensional, bounded data space.

\textbf{House Pricing Dataset}\footnote{\url{https://www.kaggle.com/c/house-prices-advanced-regression-techniques}}. This dataset utilizes predictor variables such as the number of bedrooms and lot square footage to predict house sales prices. It contains $1,460$ houses and $81$ features. For preprocessing, categorical features are converted by one-hot encoding as vectors, and each element in these vectors is regarded as an independent feature. All the features are then normalized into $275$ total features.

In the experiment, we randomly split this dataset into a training dataset and a testing dataset, with $70\%$ and $30\%$ of data, respectively, for both Fourclass and House Pricing datasets.
In addition, for each dataset, we randomly choose the corresponding proportion of the samples as the trusted clean dataset, which is then augmented to the same size as the original training dataset for the mimic model construction.

\setcounter{figure}{3}
\begin{figure*}[t]
	\centering
	\subfigure[Accuracy]{\includegraphics[scale=0.3]{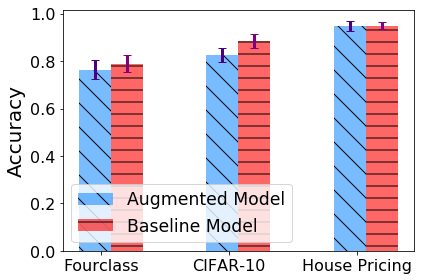}}\hspace{4mm}
	\subfigure[Recall]{\includegraphics[scale=0.3]{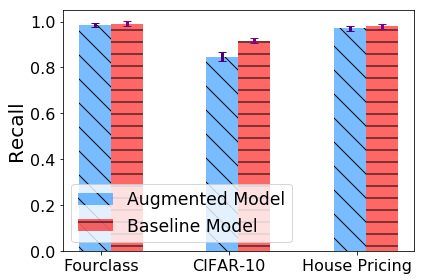}}\hspace{4mm}
	\subfigure[F1-score]{\includegraphics[scale=0.3]{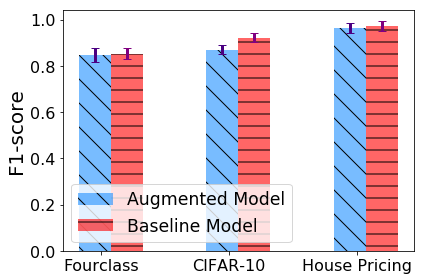}}
	\caption{Effectiveness of synthetic data generation.}
	\label{fig:comparison}
	%\vspace{0.1cm}
\end{figure*}

\subsubsection{Baselines} \label{Baselines}
We compare De-Pois to some typical defense techniques, against four types of poisoning attacks as described in Section~\ref{Related Work} and Table~\ref{tab:feasibility}.
It is noticed that existing defenses are attack-specific, and thus we only report the defending results against their corresponding attacks (i.e., those marked with {\small $\surd$} in Table~\ref{tab:feasibility}) while neglecting the ineffective defending results (i.e., those marked with $\bigcirc$ in Table~\ref{tab:feasibility}).

\subsubsection{Evaluation Metrics} \label{Metrics}
To evaluate the performance of De-Pois on defense against poisoning attacks, we adopt the standard metrics: \emph{Accuracy}, \emph{Recall} and \emph{F1-score}.
Accuracy measures the ratio of samples predicted correctly in all the predicting samples,
while F1-score is defined as the harmonic mean of \emph{Recall} and \emph{Precision}, expressed as
\begin{equation}
  F1=\frac{\textit{Precision}\times \textit{Recall}}{2\times (\textit{Precision}+\textit{Recall})},
\end{equation}
where $\textit{Recall}$ (resp. $\textit{Precision}$) measures the ratio of correctly predicted positive samples over all the positive samples (resp. all the predicted positive samples).
It is known that the higher F1-score value indicates the better performance of the defense.

To evaluate the performance of the synthetic data generation in De-Pois, we adopt the following metrics: Inception Score (IS)~\cite{salimans:2016improved}, Fr\'{e}chet Inception Distance (FID)~\cite{heusel:2017gans,dowson:1982frechet}, Wasserstein Distance (WD)~\cite{arjovsky:2017wasserstein} and Average Euclidean Distance (AED).

Specifically, IS evaluates the quality of a generated image by calculating the KL-divergence between the conditional class distribution $p(y|\boldsymbol x)$ and the marginal class distribution $p(y)$. Also, the conditional class distribution $p(y|\boldsymbol x)$ is obtained by applying the Inception model to each image. The score is given by:
\begin{equation} \begin{split}  \textit{IS}=\exp(\mathbb{E}_{\boldsymbol x \sim p_{g}}[D_{\textit{KL}}(p(y|\boldsymbol x))||p(y)]).\end{split}\end{equation}

FID measures the similarity of Gaussian distributions of Inception embedding of real and synthetic images, which is defined as:
\begin{equation}
\textit{FID}(r,g)=||\boldsymbol m_r-\boldsymbol m_g||^2+Tr(\boldsymbol C_r+\boldsymbol C_g-2(\boldsymbol C_r\boldsymbol C_g)^\frac{1}{2}),
\end{equation}
where $\boldsymbol m_r$ (resp. $\boldsymbol m_g$) and $\boldsymbol C_r$ (resp. $\boldsymbol C_g$) represent the mean and covariance of the real (resp. synthetic) image.

WD denotes the optimal transport cost in order to transform the distribution $P_r$ to $P_f$, which can be formulated as:
\begin{equation} \textit{WD}(P_r,P_f)=\inf_{\gamma \in \prod(P_r,P_f) }\mathbb{E}_{(\boldsymbol x,\boldsymbol y)\sim \gamma}\left[||\boldsymbol x-\boldsymbol y||\right],\end{equation}
where $\prod(P_r,P_f)$ is the set of joint distributions $\gamma(\boldsymbol x,\boldsymbol y)$ and $P_r,P_f$ represent the marginals of $\gamma(\boldsymbol x,\boldsymbol y)$.

AED measures the mean distance between the pair elements of $X=\left\{x_{ik}: 1\leq  i\leq M, 1\leq  k\leq K \right\}$  and $Y=\left\{y_{jk}: 1\leq  j\leq N, 1\leq  k\leq K\right\}$, expressed as
\begin{equation}
\textit{AED}=\frac{1}{M\times N}\sum_{i=1}^{M}\sum_{j=1}^{N}\left[\sum_{k=1}^{K}(x_{ik}-y_{jk})^2\right]^{\frac{1}{2}},
\end{equation}
where $M$ (resp. $N$) is the total number of samples in $X$ (resp. $Y$) and $K$ is the number of features in each sample. $x_{ik}$ (resp. $y_{jk}$) represents the value of the $k$th feature in the $i$th (resp. $j$th) sample.

Note that for each testing, we run the experiment for 5 times to average out the randomness.

\subsubsection{Model Settings} \label{Model Settings}
For synthetic data generation, we utilize cGAN architecture and choose a fixed structure to generate data.
In the generator network $G_c$, a noise prior $\boldsymbol z_c$ with $100$ dimensional normal distribution and class labels or regression values $\boldsymbol y$ of trusted dataset with a mini-batch size of $128$ (i.e., $|S_t|=20\%|S_o|$) are combined as the input at the bottommost layer of $G_c$. Also, $G_c$ use $3$ full connection layers with Leaky Rectified Linear Unit (Leaky ReLu) activation and one full connection layer with sigmoid activation. The discriminator network $D_c$ is constructed with $3$ full connection layers and a sigmoid unit layer. Dropout is applied at intermediate layers of $D_c$ and the dropout value is set to $0.4$. The slope of the leak in the Leaky ReLu is set to $0.2$ in both $G_c$ and $D_c$.

For the authenticator, we adopt a specific classifier module (i.e., CNN for images) in the context of classification, and we use the cross entropy function to calculate its loss. For regression, the authenticator calls a specific regression module (i.e., LASSO) and the loss is obtained using MSE.

For the mimic model construction, we improve WGAN-GP by feeding the class labels or regression values $\boldsymbol y$ as in cGAN into both $G_w$ and $D_w$ as additional inputs. Similarly, a noise prior $\boldsymbol z_c$ with $100$ dimensional normal distribution and class labels or regression values $\boldsymbol y$ of augmented dataset with a mini-batch size of $32$ are combined as the input at the bottommost layer of $G_w$. We use full connection layer and activating layer in both $G_w$ and $D_w$. Especially, we add convolutional layer for images in order to obtain a better mimic model. We use RMSprop optimizer with the learning rate equal to $0.00005$ and $D_w$ iterates $5$ times when $G_w$ iterates once.

\subsection{Determining Detection Boundary}

We begin with determining the detection boundary $y_{\textit{thr}}$ for each dataset using the method discussed in Section~\ref{Poisoned Data Recognition}, and we set $z_s=-1.96$ by default. Taking CIFAR-10 dataset for example, we first obtain the augmented samples consisting of $|S_t|=20\%|S_o|$ and $|S_s'|=80\%|S_o|$.
Then we calculate the mean $\mu=-0.5726$ and the standard deviation $\sigma=0.2109$ of $P_{\scriptscriptstyle S_{\textit{aug}}}$.
After that, we can compute the detection boundary by $y_{\textit{thr}}=z_s\times\sigma+\mu=-0.9859$ for CIFAR-10 dataset.

In order to evaluate our detection boundary, we test $12,000$ clean samples and $12,000$ poisoned samples generated by TCL-attack for CIFAR-10 dataset. Then the prediction values of testing samples are calculated after being fed into the detection model trained on the augmented samples. The distribution of the prediction values are depicted in Fig.~\ref{fig:mnist_threshold}. We can observe that our detection boundary is very close to the ideal one and the poisoned sample can be distinguished from the clean sample given our detection boundary. Similarly, we determine the detection boundary $y_{\textit{thr}}=-0.073$ for Fourclass dataset and $y_{\textit{thr}}=0.683$ for House Pricing dataset.
In the following, we evaluate the performance of De-Pois using the obtained detection boundary for each dataset.

\setcounter{figure}{2}
\begin{figure}[t]
	\centering
	\includegraphics[scale=0.4]{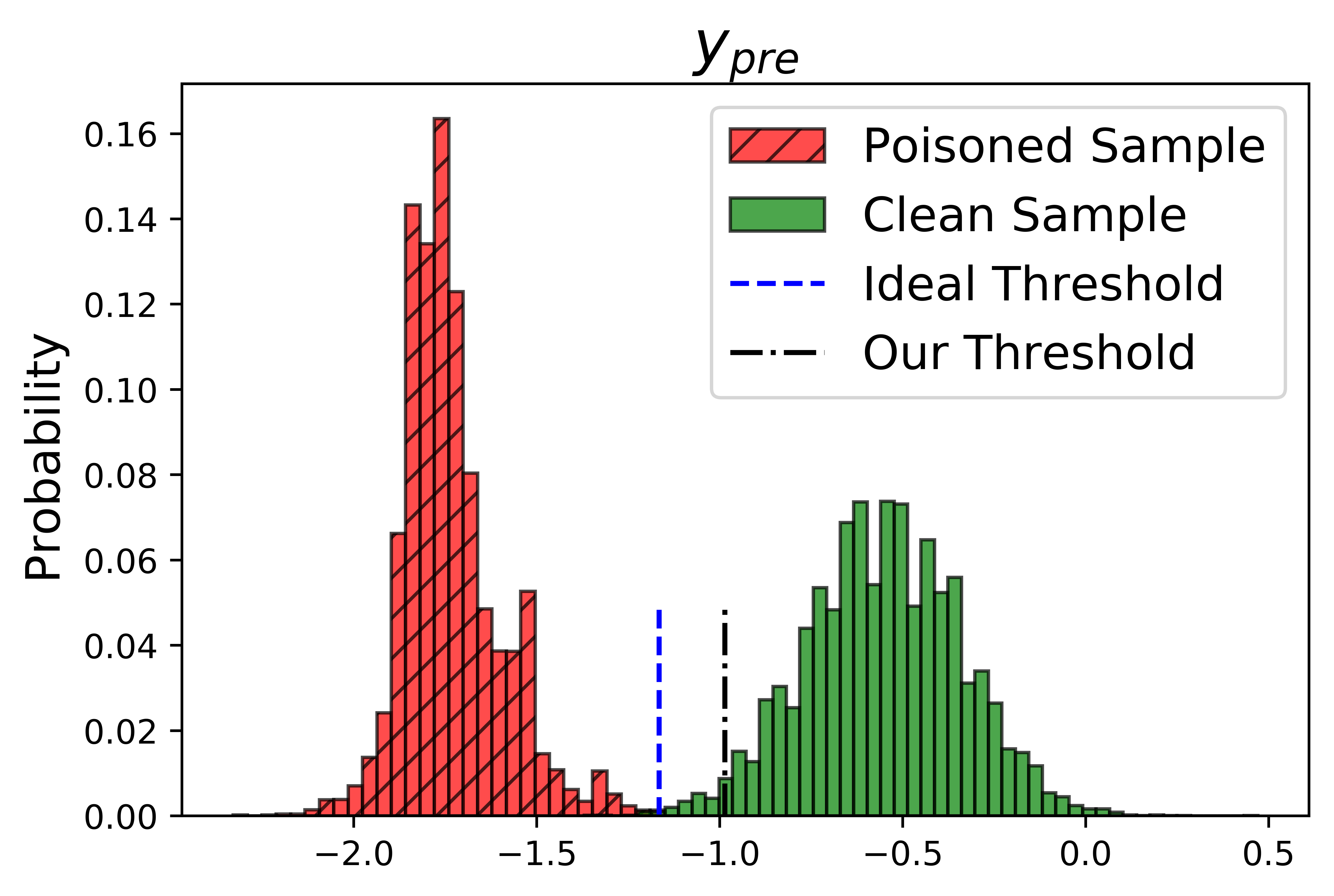}
	\caption{Distributions of prediction values of the clean samples and the poisoned samples generated by TCL-attack.}
	\label{fig:mnist_threshold}
	%\vspace{0.1cm}
\end{figure}

\setcounter{figure}{4}
\begin{figure*}[t]
	\centering
	\subfigure[TCL-attack]{\label{fig:a}\includegraphics[width=37mm]{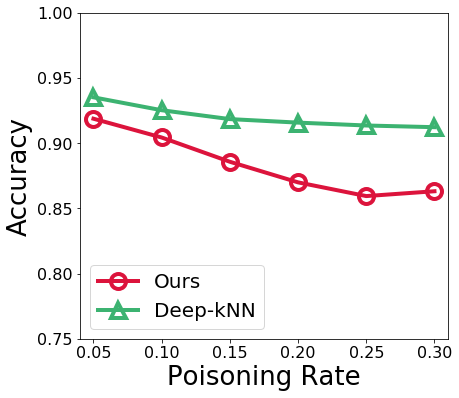}}\hspace{4mm}
	\subfigure[pGAN-attack]{\label{fig:b}\includegraphics[width=37mm]{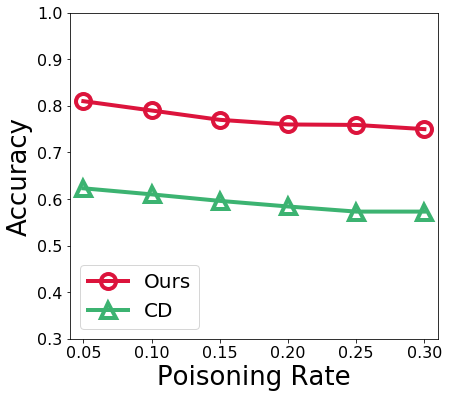}}\hspace{4mm}
	\subfigure[LF-attack]{\label{fig:c}\includegraphics[width=37mm]{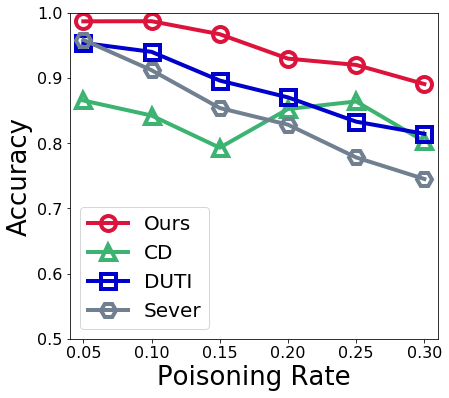}}\hspace{4mm}
	\subfigure[R-attack]{\label{fig:d}\includegraphics[width=37mm]{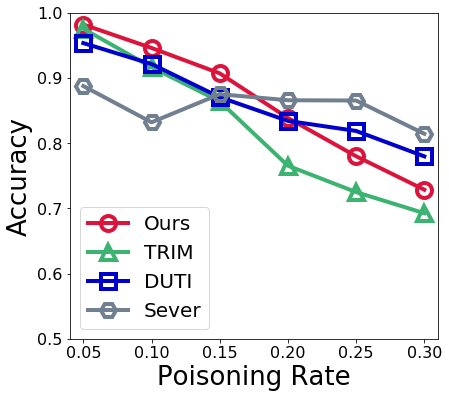}}
	\vspace{-4pt}
	\caption{Impact of poisoning rate on accuracy in fixed trusted clean set environments.}
	\label{fig:poisoning_rate}\vspace{-4pt}
\end{figure*}

\begin{figure*}[t]
	\centering \subfigure[TCL-attack]{\label{fig:a}\includegraphics[width=37mm]{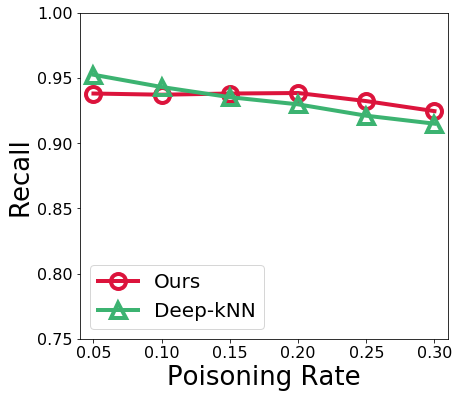}}\hspace{4mm} \subfigure[pGAN-attack]{\label{fig:b}\includegraphics[width=37mm]{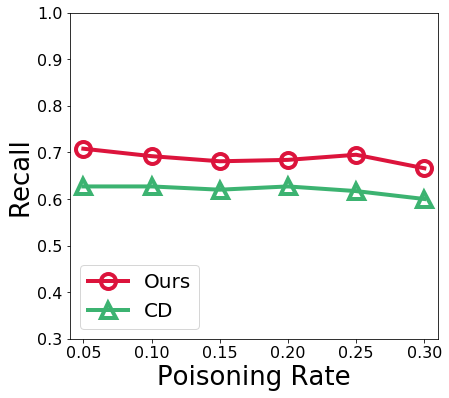}}\hspace{4mm}
	\subfigure[LF-attack]{\label{fig:c}\includegraphics[width=37mm]{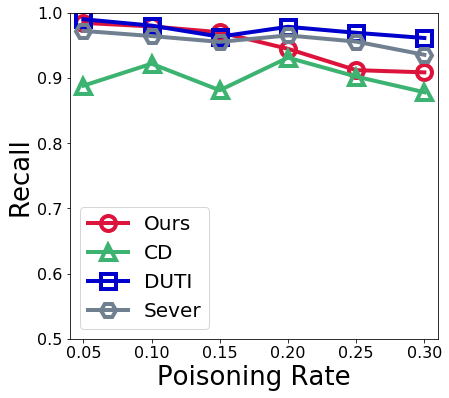}}\hspace{4mm} \subfigure[R-attack]{\label{fig:d}\includegraphics[width=37mm]{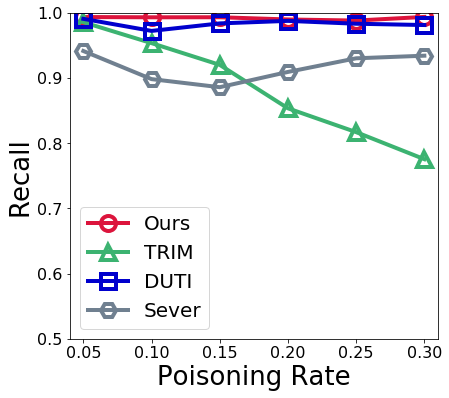}}
	\vspace{-4pt}
	\caption{Impact of poisoning rate on recall in fixed trusted clean set environments.}
	\label{fig:trusteS_trainingset_recall}\vspace{-4pt}
\end{figure*}

\begin{figure*}[t]
	\centering
	\subfigure[TCL-attack]{\label{fig:a}\includegraphics[width=37mm]{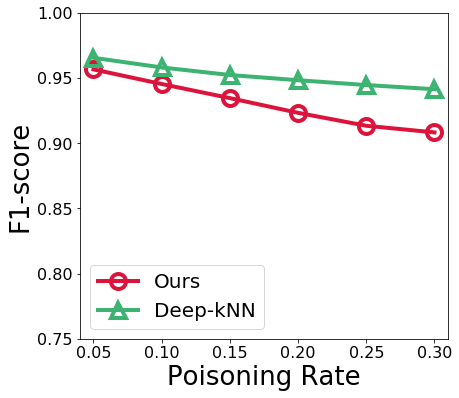}}\hspace{4mm}
	\subfigure[pGAN-attack]{\label{fig:b}\includegraphics[width=37mm]{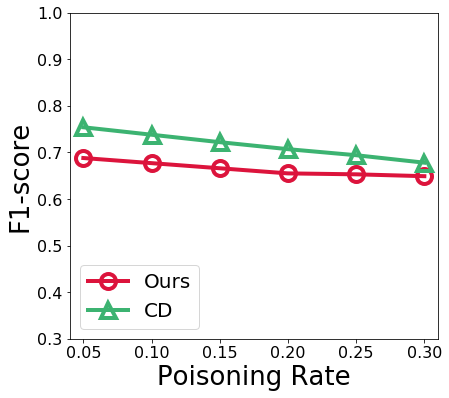}}\hspace{4mm}
	\subfigure[LF-attack]{\label{fig:c}\includegraphics[width=37mm]{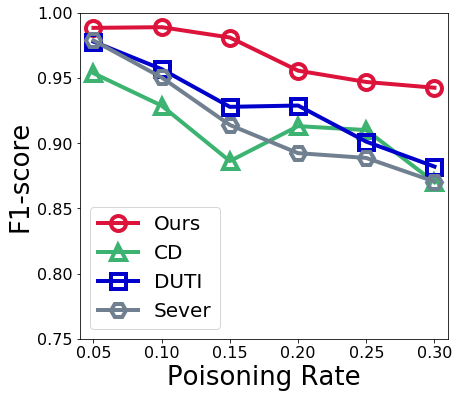}}\hspace{4mm}
	\subfigure[R-attack]{\label{fig:d}\includegraphics[width=37mm]{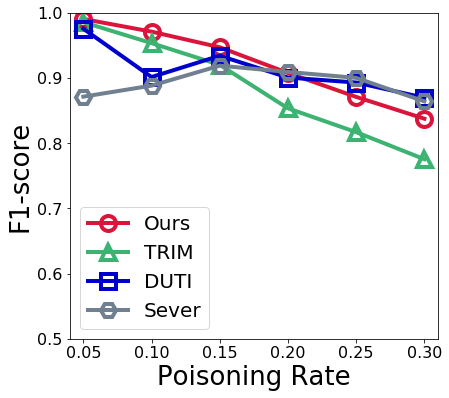}}
	\vspace{-4pt}
	\caption{Impact of poisoning rate on F1-score in fixed trusted clean set environments.}
	\label{fig:trusteS_trainingset}\vspace{-4pt}
\end{figure*}

\subsection{Effectiveness of Synthetic Data Generation}
We first evaluate the effectiveness of the synthetic data generation, by comparing the accuracy, recall and F1-score of the two models: the {\em augmented model} trained on $S_{\textit{aug}}$, and the {\em baseline model} trained on clean dataset with the same size as $S_{\textit{aug}}$.
The two models are trained with our cGAN and authenticator architecture on each dataset, and the trained discriminator is employed to evaluate the performance.
To unify metrics for both classification and regression datasets, we test the two models on data with both clean and poisoned samples. For the regression dataset, we can thus obtain the results passing the model and then calculate the accuracy, the recall and F1-score as for the classification dataset.

For CIFAR-10, Fourclass and House Pricing datasets, we test $15,000$ clean data and $15,000$ poisoned data generated by TCL-attack, $500$ clean data and $500$ poisoned data generated by LF-attack, and $1,000$ clean data and $1,000$ poisoned data generated by R-attack, respectively. From Fig.~\ref{fig:comparison} we can see that, the accuracy, recall and F1-score of the augmented model are quite close to those of the baseline model, thereby indicating that our synthetic data generation is effective for augmenting the trusted clean data.

Furthermore, Table~\ref{tab:different metric on three dataset} shows the IS and FID results on CIFAR-10 dataset, as well as WD and AED results on Fourclass and House Pricing datasets obtained by our synthetic data generation model and a comparison of baseline on real data and cGAN model.
We can observe that our model can reach the performance as that from real data and outperforms the cGAN model, indicating that using authenticator could be beneficial to generate sufficient valid data.

\subsection{Impact of Poisoning Rate}
We next evaluate the impact of poisoning rate $R_p=|S_p|/|S_c|$.
In our settings, we set $|S_t|=20\%|S_o|$.
For each type of poisoning attacks, the results are reported in Figs.~\ref{fig:poisoning_rate}$\sim$\ref{fig:trusteS_trainingset}, and it is confirmed that De-Pois is attack-agnostic, while other defenses are attack-specific.
We describe the details as follows.

\begin{table}[t]
	\caption{Evaluation of synthetic data generation.}  \label{tab:different metric on three dataset}
	{\fontsize{9}{11}\selectfont
		\begin{tabular}
			{|c|p{0.65cm}<{\centering} |c|p{0.6cm}|c|p{0.65cm}|c|}
			\hline
			\multicolumn{1}{|c|}{Dataset} & \multicolumn{2}{c|}{CIFAR-10}                 & \multicolumn{2}{c|}{Fourclass}                 & \multicolumn{2}{c|}{House Pricing}                 \\ \hline
			\diagbox[width=6em,trim=l]{Model}{Metric}                      & IS               & FID             & WD              & AED              & WD               & AED             \\ \hline
			Real data         & 11.33           & 2.1            & 0.00             & 87.38             & 0.00              & 5.01                \\ \hline
			cGAN              & 6.10           & 51.03            & 0.89             & 91.65             & 2.20              & 7.93               \\ \hline
			De-Pois              & 8.23           & 15.2            & 0.61             & 87.97             & 1.98              &   7.84               \\ \hline
		\end{tabular}
		
		\begin{tablenotes}
			\item[]{\small IS: higher is better. \qquad FID\,/\,WD\,/\,AED: lower is better. }
		\end{tablenotes}
		
	}
\end{table}

\setcounter{figure}{7}
\begin{figure*}[t]
	\centering
	\subfigure[TCL-attack]{\label{fig:each part:a}\includegraphics[width=42mm]{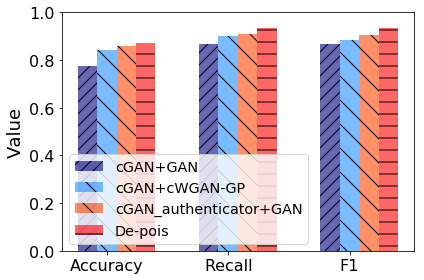}}\hspace{3pt}
	\subfigure[pGAN-attack]{\label{fig:each part:b}\includegraphics[width=42mm]{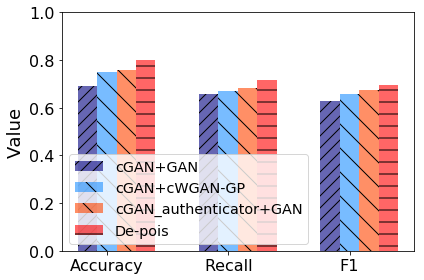}}\hspace{3pt}
	\subfigure[LF-attack]{\label{fig:each part:c}\includegraphics[width=42mm]{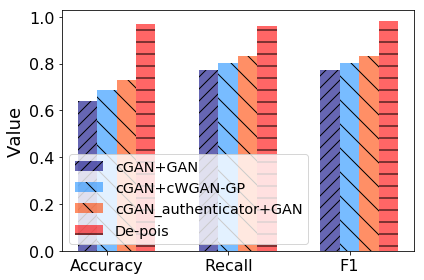}}\hspace{3pt}
	\subfigure[R-attack]{\label{fig:each part:d}\includegraphics[width=42mm]{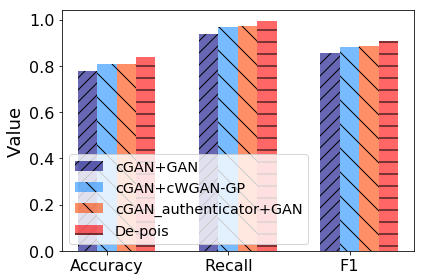}}
	\vspace{-3pt}
	\caption{Impact of each part in De-Pois.}
	\label{fig:each part}\vspace{-5pt}
\end{figure*}

\subsubsection{Effectiveness under TCL-attack on CIFAR-10 Dataset}
We first assess De-Pois against TCL-attack~\cite{zhu:2019transferable}. We use CIFAR-10 dataset and select ``ship'' as the target class to craft the poisons, and ``frog'' as the targeted class. We compare De-Pois with Deep-$k$NN. From the results shown in Figs.~\ref{fig:poisoning_rate}$\sim$\ref{fig:trusteS_trainingset}, we observe that Deep-$k$NN maintains relatively higher accuracy, recall and F1-score. This is largely because Deep-$k$NN defense against this kind of feature collision attack in feature space and find the characteristic that the plurality of poisoned data's neighbors as clean ones in the target class.
It is also noted that although the accuracy, recall and F1-score of De-Pois is over $3\%$ lower than Deep-$k$NN on average, De-Pois still works in general, with the accuracy over $0.85$ and F1-score over $0.9$ in most cases.

\subsubsection{Effectiveness under pGAN-attack on CIFAR-10 Dataset}
We then evaluate De-Pois against pGAN-attack~\cite{munoz:2019poisoning}. We use CIFAR-10 dataset and set $|S_t|=20\%|S_o|$.
We consider detectability constraints in pGAN-attack in realistic scenarios and observe that pGAN-attack can produce effective poisoning data when $\alpha=0.5$.
If $\alpha$ is small, the performance of De-Pois will be better because pGAN-attack considers less detectability constraint then; when the value of $\alpha$ increases, the generated poisoning data will become hard to detect for both CD and De-Pois.
Since $\alpha$ controls the stealth of the attack and, as a trade-off, $\alpha = 0.5$ is set in our experiment.
We compare De-Pois with CD, and the results are shown in Figs.~\ref{fig:poisoning_rate}$\sim$\ref{fig:trusteS_trainingset}.
As can be observed, the accuracy and recall of De-Pois are always higher than CD when the poisoning data increases, with an average $15\%$ and $7\%$ improvement, respectively. It is also shown that the F1-score of CD is 3\% higher on average than that of De-Pois. The results indicate that De-Pois can defense pGAN-attack well.

\begin{table*}[t]
	\centering
	\caption{Comparison of accuracy and F1-score in fixed poisoning rate environments.}  %表格标题
	{\fontsize{8.5}{11.5}\selectfont
		%	\footnotesize
		\setlength{\tabcolsep}{1.2mm}{
			\begin{tabular}{cllllllll} %表格8列 全部居中显示
				\toprule
				&        &        & \multicolumn{6}{c}{$\bm{|S_t|}$} \\
				\cline{4-9}
				\multicolumn{1}{l}{\textbf{Metric}} & \textbf{Attack} & \textbf{Defense} & \multicolumn{1}{c}{5\%$|S_o|$} & \multicolumn{1}{c}{10\%$|S_o|$} & \multicolumn{1}{c}{15\%$|S_o|$} & \multicolumn{1}{c}{20\%$|S_o|$} & \multicolumn{1}{c}{25\%$|S_o|$} & \multicolumn{1}{c}{30\%$|S_o|$} \\
				\cline{1-9}  %为第1列到第8列添加横线
				\hline \hline
				
			&\multicolumn{1}{l}{\multirow{2}{*}{TCL-attack}}&Ours&0.756$\pm$0.05&0.843$\pm$0.05&0.848$\pm$0.03&0.869$\pm$0.03&0.851$\pm$0.03&0.927$\pm$0.02 \\			\cline{3-9}
				&&\textbf{Deep-$k$NN}&\textbf{0.915$\pm$0.02}&\textbf{0.905$\pm$0.02}&\textbf{0.910$\pm$0.02}&\textbf{0.909$\pm$0.02}&\textbf{0.918$\pm$0.02}&\textbf{0.912$\pm$0.02} \\
				
				\cline{2-9}  %为第2列到第8列添加横线
				
				&\multicolumn{1}{c}{\multirow{2}{*}{pGAN-attack}}&\textbf{Ours}&\textbf{0.756$\pm$0.03}&\textbf{0.767$\pm$0.03}&\textbf{0.781$\pm$0.03}&\textbf{0.783$\pm$0.02}&\textbf{0.790$\pm$0.02}&\textbf{0.828$\pm$0.02} \\
				\cline{3-9}
				&&CD&0.635$\pm$0.03&0.631$\pm$0.03&0.626$\pm$0.03&0.636$\pm$0.03&0.631$\pm$0.03&0.632$\pm$0.03 \\
				
				\cline{2-9}  %为第2列到第8列添加横线
				&&\textbf{Ours}&\textbf{0.889$\pm$0.05}&\textbf{0.919$\pm$0.05}&\textbf{0.922$\pm$0.04}&\textbf{0.929$\pm$0.04}&\textbf{0.933$\pm$0.04}&\textbf{0.931$\pm$0.03} \\
				\cline{3-9}
				Accuracy 	&LF-attack&CD&0.736$\pm$0.03&0.745$\pm$0.03&0.762$\pm$0.03&0.769$\pm$0.03&0.771$\pm$0.02&0.782$\pm$0.02 \\
				\cline{3-9}
				&&DUTI&0.766$\pm$0.04&0.785$\pm$0.04&0.809$\pm$0.04&0.821$\pm$0.04&0.853$\pm$0.03&0.873$\pm$0.03 \\
				\cline{3-9}
				&&Sever&0.845$\pm$0.04&0.857$\pm$0.03&0.852$\pm$0.03&0.840$\pm$0.04&0.848$\pm$0.04&0.837$\pm$0.04 \\
				
				\cline{2-9}  %为第2列到第8列添加横线
				&&Ours&0.832$\pm$0.03&0.836$\pm$0.03&0.839$\pm$0.03&0.838$\pm$0.03&0.836$\pm$0.03&0.840$\pm$0.03 \\
				\cline{3-9}
				&R-attack&TRIM&0.772$\pm$0.05&0.775$\pm$0.05&0.771$\pm$0.05&0.768$\pm$0.05&0.772$\pm$0.05&0.775$\pm$0.05 \\
				\cline{3-9}
				&&DUTI&0.885$\pm$0.02&0.890$\pm$0.02&0.890$\pm$0.02&0.886$\pm$0.02&0.885$\pm$0.02&0.890$\pm$0.02 \\
				\cline{3-9}
				&&\textbf{Sever}&\textbf{0.856$\pm$0.03}&\textbf{0.865$\pm$0.02}&\textbf{0.860$\pm$0.02}&\textbf{0.863$\pm$0.02}&\textbf{0.848$\pm$0.02}&\textbf{0.852$\pm$0.02} \\
				
				\cline{1-9}  %%%%%%%%%%%%%%%%%为第1列到第8列添加横线 recall
				\hline \hline
				
				&\multicolumn{1}{l}{\multirow{2}{*}{TCL-attack}}&Ours&0.744$\pm$0.05&0.887$\pm$0.05&0.867$\pm$0.03&\textbf{0.938$\pm$0.03}&0.873$\pm$0.03&\textbf{0.967$\pm$0.01} \\
				\cline{3-9}
				&&\textbf{Deep-$k$NN}&\textbf{0.927$\pm$0.02}&\textbf{0.913$\pm$0.03}&\textbf{0.925$\pm$0.02}&0.925$\pm$0.02&\textbf{0.935$\pm$0.03}&0.930$\pm$0.03 \\

				\cline{2-9}  %为第2列到第8列添加横线
				&\multicolumn{1}{c}{\multirow{2}{*}{pGAN-attack}}	&\textbf{Ours}&\textbf{0.674$\pm$0.04}&\textbf{0.688$\pm$0.03}&\textbf{0.702$\pm$0.04}&\textbf{0.711$\pm$0.03}&\textbf{0.722$\pm$0.03}&\textbf{0.735$\pm$0.03} \\
				\cline{3-9}
			    &&CD&0.638$\pm$0.02&0.641$\pm$0.02&0.638$\pm$0.02&0.636$\pm$0.02&0.631$\pm$0.02&0.640$\pm$0.02 \\
			
				\cline{2-9}  %为第2列到第8列添加横线
				&&\textbf{Ours}&0.930$\pm$0.04&\textbf{0.943$\pm$0.04}&\textbf{0.950$\pm$0.04}&\textbf{0.954$\pm$0.04}&\textbf{0.961$\pm$0.04}&\textbf{0.958$\pm$0.04} \\
				\cline{3-9}
				Recall	&LF-attack&CD&0.926$\pm$0.03&0.931$\pm$0.03&0.932$\pm$0.03&0.937$\pm$0.03&0.916$\pm$0.03&0.922$\pm$0.03 \\
				\cline{3-9}
				&&DUTI&0.916$\pm$0.02&0.922$\pm$0.02&0.934$\pm$0.02&0.938$\pm$0.02&0.947$\pm$0.02&0.953$\pm$0.02 \\
				\cline{3-9}
				&&Sever&\textbf{0.941$\pm$0.03}&0.955$\pm$0.02&0.950$\pm$0.02&0.939$\pm$0.03&0.947$\pm$0.03&0.936$\pm$0.03 \\
				
				\cline{2-9}  %为第2列到第8列添加横线
				&&\textbf{Ours}&\textbf{0.992$\pm$0.03}&\textbf{0.985$\pm$0.03}&\textbf{0.967$\pm$0.03}&\textbf{0.980$\pm$0.03}&\textbf{0.988$\pm$0.02}&\textbf{0.973$\pm$0.02} \\
				\cline{3-9}
				&R-attack&TRIM&0.857$\pm$0.05&0.863$\pm$0.05&0.855$\pm$0.05&0.861$\pm$0.05&0.864$\pm$0.05&0.885$\pm$0.05 \\
				\cline{3-9}
				&&DUTI&0.953$\pm$0.02&0.960$\pm$0.02&0.960$\pm$0.02&0.957$\pm$0.02&0.955$\pm$0.02&0.955$\pm$0.02 \\
				\cline{3-9}
				&&Sever&0.893$\pm$0.03&0.912$\pm$0.02&0.903$\pm$0.03&0.908$\pm$0.03&0.889$\pm$0.03&0.890$\pm$0.03 \\
				
				\cline{1-9}  %%%%%%%%%%%%%%%%为第1列到第8列添加横线  f1
				\hline \hline
				
				&\multicolumn{1}{l}{\multirow{2}{*}{TCL-attack}}&Ours&0.851$\pm$0.05&0.904$\pm$0.05&0.903$\pm$0.03&0.923$\pm$0.03&0.892$\pm$0.03&\textbf{0.956$\pm$0.02} \\
				\cline{3-9}
				&&\textbf{Deep-$k$NN}&\textbf{0.948$\pm$0.02}&\textbf{0.933$\pm$0.02}&\textbf{0.943$\pm$0.03}&\textbf{0.939$\pm$0.03}&\textbf{0.952$\pm$0.02}&0.945$\pm$0.02 \\

				\cline{2-9}  %为第2列到第8列添加横线
				&\multicolumn{1}{c}{\multirow{2}{*}{pGAN-attack}}	&\textbf{Ours}&0.678$\pm$0.03&0.685$\pm$0.03&0.700$\pm$0.03&0.703$\pm$0.03&\textbf{0.713$\pm$0.03}&0.722$\pm$0.03 \\
				\cline{3-9}
				&&CD&\textbf{0.725$\pm$0.02}&\textbf{0.713$\pm$0.02}&\textbf{0.721$\pm$0.02}&\textbf{0.718$\pm$0.02}&0.711$\pm$0.02&\textbf{0.724$\pm$0.02} \\
				
				\cline{2-9}  %为第2列到第8列添加横线
				&&\textbf{Ours}&\textbf{0.930$\pm$0.05}&\textbf{0.943$\pm$0.05}&\textbf{0.950$\pm$0.05}&\textbf{0.954$\pm$0.05}&\textbf{0.961$\pm$0.05}&\textbf{0.958$\pm$0.05} \\
				\cline{3-9}
				F1-score	&LF-attack&CD&0.826$\pm$0.02&0.828$\pm$0.02&0.832$\pm$0.02&0.837$\pm$0.02&0.842$\pm$0.01&0.844$\pm$0.01 \\
				\cline{3-9}
				&&DUTI&0.871$\pm$0.03&0.883$\pm$0.03&0.898$\pm$0.03&0.911$\pm$0.03&0.924$\pm$0.02&0.935$\pm$0.02 \\
				\cline{3-9}
				&&Sever&0.885$\pm$0.03&0.905$\pm$0.02&0.897$\pm$0.02&0.876$\pm$0.03&0.882$\pm$0.03&0.869$\pm$0.03 \\
				
				\cline{2-9}  %为第2列到第8列添加横线
				&&Ours&0.907$\pm$0.03&0.907$\pm$0.03&0.908$\pm$0.03&0.910$\pm$0.03&\textbf{0.908$\pm$0.03}&0.906$\pm$0.03 \\
				\cline{3-9}
				&R-attack&TRIM&0.858$\pm$0.05&0.880$\pm$0.05&0.882$\pm$0.05&0.885$\pm$0.05&0.886$\pm$0.05&0.885$\pm$0.05 \\
				\cline{3-9}
				&&DUTI&0.895$\pm$0.02&0.899$\pm$0.02&0.895$\pm$0.02&0.90$\pm$0.02&0.897$\pm$0.02&0.90$\pm$0.02 \\
				\cline{3-9}
				&&\textbf{Sever}&\textbf{0.912$\pm$0.03}&\textbf{0.923$\pm$0.02}&\textbf{0.915$\pm$0.03}&\textbf{0.918$\pm$0.03}&0.901$\pm$0.02&\textbf{0.906$\pm$0.02} \\
				\bottomrule
		\end{tabular}}
	}
	\label{tab:trainingset}
\end{table*}

\subsubsection{Effectiveness under LF-attack on Fourclass Dataset}
We next assess De-Pois under LF-attack strategy~\cite{Biggio:2011SupportVM}.
We use Fourclass dataset and set $|S_t|=20\%|S_o|$.
We compare De-Pois with DUTI, Sever and CD. The results in Figs.~\ref{fig:poisoning_rate}$\sim$\ref{fig:trusteS_trainingset} illustrate that De-Pois always performs better than other three methods as $R_p$ increases, as De-Pois is good at learning the distribution of features especially for classification tasks. Also, DUTI has comparable performance with Sever. The accuracy and F1-score of DUTI (resp. Sever) is over 0.85 and 0.85 (resp. 0.8 and 0.85), respectively, as $R_p$ increases.

\subsubsection{Effectiveness under R-attack on House Pricing Dataset}
Finally, we analyze De-Pois against R-attack~\cite{Jagielski:2018ManipulatingML}.
We use House Pricing dataset and set $|S_t|=20\%|S_o|$.
We compare De-Pois with TRIM, DUTI and Sever.
In Figs.~\ref{fig:poisoning_rate}$\sim$\ref{fig:trusteS_trainingset}, the results indicate that De-Pois always performs well than TRIM. The main reason is that TRIM identifies the poisoned data with the lowest residual. In contrast, De-Pois can identify poisoned samples whose prediction values of the mimic model are lower than those of the clean ones.
Moreover, De-Pois performs better than Sever and DUTI when the poisoned rate is lower than $20\%$, while Sever and DUTI perform better when the poisoned rate continues increasing. This is largely because Sever and DUTI contain several retrain process after they detect outliers and thus have more chance to filter out the possible poisoned samples.
It can also be seen that De-Pois performs well regardless of the way how poisoning samples are manipulated, which makes De-Pois a more generic solution.

\subsection{Impact of Trusted Clean Data Size}\label{Impact of Trusted Clean Set Size}
To evaluate the impact of the size of $S_t$, we set $R_p$ to a fixed value (e.g. $R_p=20\%$ in our settings), and conduct experiments in a similar way to the previous section.
The results are shown in Table~\ref{tab:trainingset}, which reveal that the accuracy, recall and F1-score of De-Pois are higher than $0.9$ on average and fluctuate less than $10\%$ as the size of $S_t$ increases under TCL-attack, LF-attack and R-attack. It is noticed that De-Pois performs well against pGAN-attack with rational detectability constraints.
Also, it has comparable effectiveness with Deep-$k$NN against TCL-attack, even when $|S_t|=5\%|S_o|$, which verifies again that De-Pois is attack-agnostic, and is effective for both classification and regression tasks.

\subsection{Impact of Each Component in De-Pois}
We also evaluate the impact of each component in De-Pois.
We compare De-Pois (i.e. cGAN\_authenticator$+$cWGAN-GP, representing cGAN with the authenticator for synthetic data generation, and cWGAN-GP for mimic model construction) with three other different compositions, namely cGAN$+$GAN, cGAN$+$cWGAN-GP and cGAN\_authenticator$+$GAN.
In the experiments $R_p=20\%$ and $|S_t|=20\%|S_o|$.
The results are shown in Fig.~\ref{fig:each part}, our authenticator improves on accuracy, recall, and F1 by at least $0.03$, $0.04$ and $0.03$, respectively.
Also, our mimic model construction part improves on accuracy, recall, and F1 by at least $0.05$, $0.04$ and $0.03$, respectively.
In a word, De-Pois demonstrates the effectiveness and performance improvements of all the components used simultaneously.

\subsection{Effectiveness under GP-attack on MNIST Dataset} \label{GP-attack_experiment}
We then evaluate De-Pois against GP-attack developed in~\cite{Yang:2017GenerativePA}.
We apply the direct gradient method to craft poisoned training data.
We also use the MNIST dataset and set $|S_t|=20\%|S_o|$.
We compare De-Pois with CD.
As can be observed in Fig.~\ref{fig:GP-attack}, the accuracy and F1-score of De-Pois are always higher than CD when $R_p$ increases, with an average $13\%$ and $10\%$ improvement, respectively. The primary cause is that CD was initially designed for LF-attack, and thus is not that well-directed for GP-attack.
mainly due to the high sensitivity of the presence of the poisoned data.

\setcounter{figure}{8}
\begin{figure}[t]
	\centering	\subfigure[Accuracy]{\includegraphics[scale=0.23]{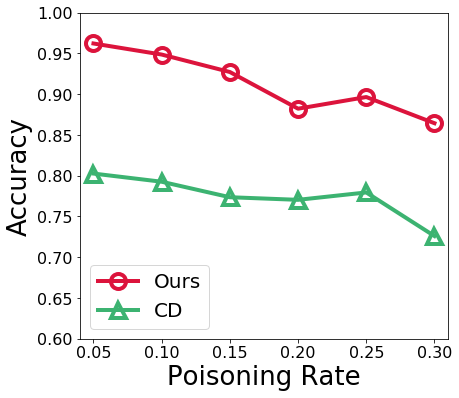}}\hspace{1mm}
	\subfigure[F1-score]{\includegraphics[scale=0.23]{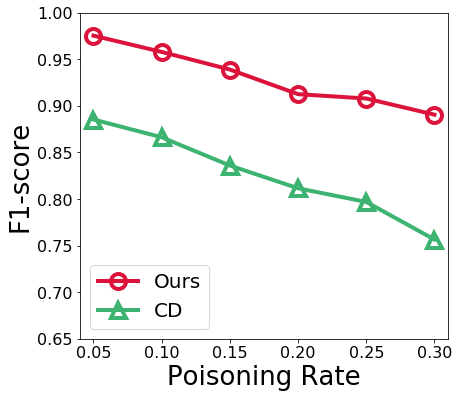}}
	\vspace{-4pt}
	\caption{Effectiveness under GP-attack on MNIST dataset.}
	\label{fig:GP-attack}
	%\vspace{0.1cm}
\end{figure}

\subsection{Evaluation on Limitations of De-Pois} \label{limitations_experiment}
Since De-Pois largely relies on the small portion of the trusted clean data, it is thus crucial to explore the impact of the trusted clean data regarding the limitations of De-Pois.

Intuitively, different sizes of the trusted clean training dataset could result in different prediction performances. It is thus reasonable to wonder that whether De-Pois is still applicable even when the amount of trust dataset is extremely small.
We push the trusted data ratio to the extreme to test where De-Pois would fail. We set $|S_t|$ to $0.5\%|S_o|$, $1\%|S_o|$, $2\%|S_o|$, $3\%|S_o|$, $4\%|S_o|$, $5\%|S_o|$ and test TCL-attack on CIFAR-10 dataset, LF-attack on Fourclass dataset, and R-attack on House Pricing dataset. We observe from Fig.~\ref{fig:limitation_one} that the accuracy and F1-score decrease when the trusted data ratio decreases.
The accuracy and F1-score decrease from $90\%$ to $70\%$ when $|S_t|$ decreases from $5\%|S_o|$ to $0.5\%|S_o|$ on average. This illustrates that the trusted data ratio could have a great impact on the performance of De-Pois and it performs not so well when the trusted data ratio is quite low.

In practice, obtaining $100\%$ clean data may be difficult, and it is sometimes inevitable that there could exist a small percentage of poisoned samples in the trusted data.
We thus alter the ratio of poisoned samples in the trusted data, and test TCL-attack on CIFAR-10 dataset, LF-attack on Fourclass dataset, and R-attack on House Pricing dataset to see how De-Pois works in this situation. As illustrated in Fig.~\ref{fig:limitation_two}, we notice that the accuracy and F1-score for three datasets decrease 10\% on average when there exist poisoned data in the trusted training dataset. Also, the accuracy and F1-score fluctuate less than 5\% when the poisoned data ratio continue increases. It should be emphasized that the performance of De-Pois drops fast when the poisoned data ratio is higher than 20\%. This illustrates that the performance of De-Pois would be impacted when there is poisoned data mixed in the trusted data.

\setcounter{figure}{9}
\begin{figure}[t]
	\centering
	\subfigure[Accuracy]{\includegraphics[scale=0.23]{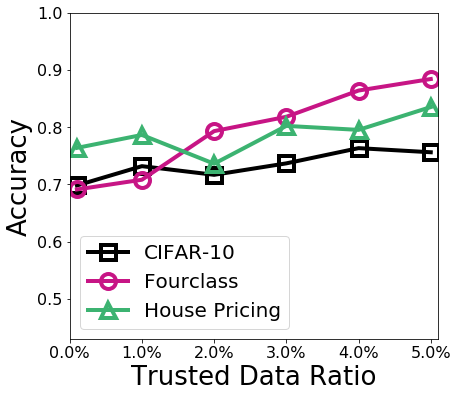}}\hspace{1mm}
 	\subfigure[F1-score]{\includegraphics[scale=0.23]{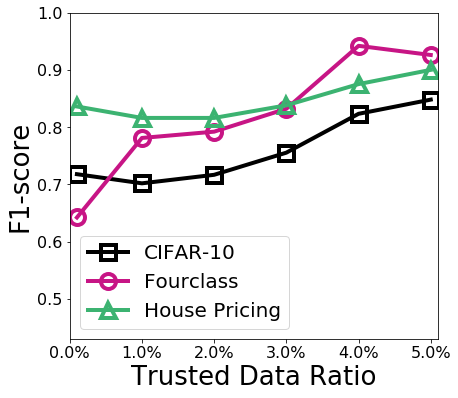}}
	\vspace{-4pt}
    \caption{Performance with insufficient trusted data.}
	\label{fig:limitation_one}
	%\vspace{0.1cm}
\end{figure}

\setcounter{figure}{10}
\begin{figure}[t]
	\centering
	\subfigure[Accuracy]{\includegraphics[scale=0.23]{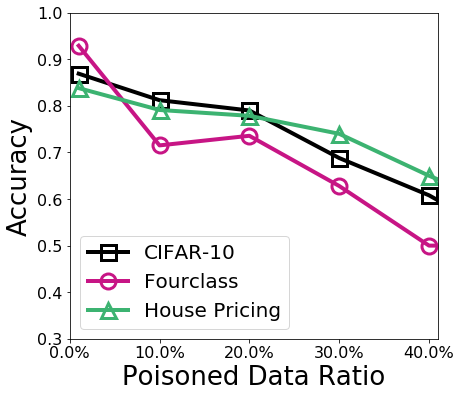}}\hspace{2mm}
	\subfigure[F1-score]{\includegraphics[scale=0.23]{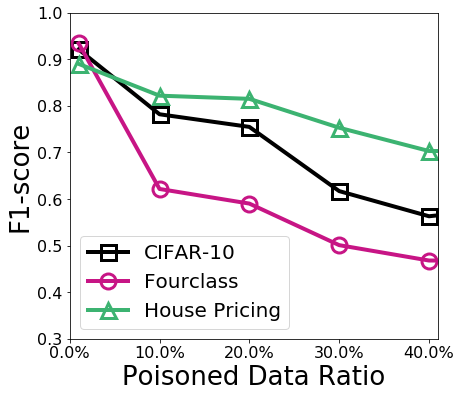}}
	\vspace{-4pt}
    \caption{Performance with poisoned data mixed in trusted data.}
	\label{fig:limitation_two}
	%\vspace{0.1cm}
\end{figure}

\setcounter{figure}{11}
\begin{figure}[t]
	\centering
	\subfigure[]{\includegraphics[scale=0.23]{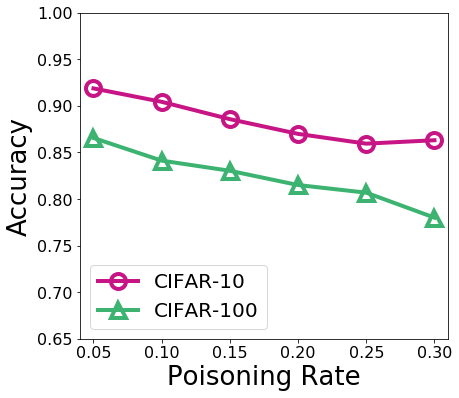}}\hspace{2mm}
	\subfigure[]{\includegraphics[scale=0.23]{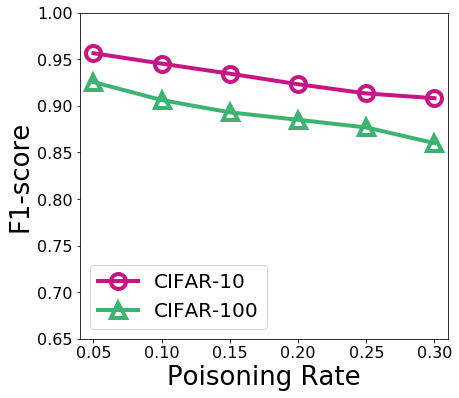}}
	\vspace{-4pt}
	\caption{Performance on CIFAR-100 dataset.}
	\label{fig:limitation_three}
	%\vspace{0.1cm}
\end{figure}

\setcounter{figure}{12}
\begin{figure}[t]
	\centering
	\includegraphics[scale=0.25]{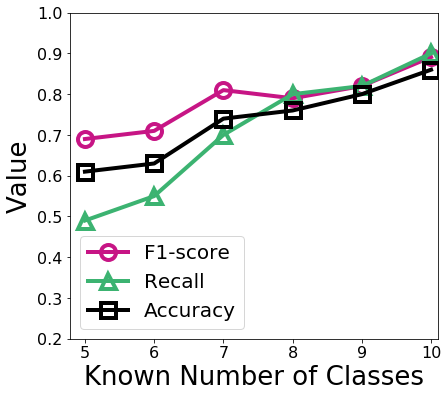}
	\caption{Performance of different known number of classes.}
	\label{fig:limitation_four}
	%\vspace{0.1cm}
\end{figure}

In addition, complex datasets with more classes could cause the performance degradation of De-Pois. We test De-Pois on CIFAR-100 dataset under TCL-attack and compare the results with those on CIFAR-10 dataset under the same experimental settings. We set $|S_t|$=$20\%|S_o|$ and alter the poisoning rate from 5\% to 30\%. We can observe from Fig.~\ref{fig:limitation_three} that the accuracy and F1-score of CIFAR-10 are 6\% and 3\% higher than those of CIFAR-100 on average. This illustrates that the performance of De-Pois will degrade on some more complex datasets.

Moreover, the synthetic data generation of De-Pois aims to enlarge the training dataset, but it depends on annotated datasets and would become less effective to generate synthetic training samples with labels not belonging to the category of these annotated datasets. In other words, our data augmentation method implicitly assumes to obtain the training samples of every class. We obtain the trusted samples from part of $10$ classes with CIFAR-10 under TCL-attack, and randomly select the known number of classes in CIFAR-10 dataset (e.g., from $5$ classes to $10$ classes). Also, we set the trusted data $|S_t|$=$20\%|S_o|$. Then we test the recall, accuracy and F1-score on $10,000$ clean samples and $10,000$ poisoned samples. The results in Fig.~\ref{fig:limitation_four} have shown that the recall, accuracy and F1-score indeed decrease when the number of classes of trusted dataset decreases.

\subsection{Evaluation on Run-time Overhead} \label{Detection Time Overhead}
Lastly, we evaluate De-Pois' run-time overhead, and compare with other defenses.
We ran our experiments on NVIDIA Geforce GTX $1060$ $6$GB and the experimental results are shown in Tables~\ref{tab:detection_time} and~\ref{tab:time overhead}. The detection time for De-Pois is calculated after we obtain the mimic model and we count the total time for detection $1,000$ samples.
 For CIFAR-10 dataset, De-Pois takes $0.108$ seconds for detecting $1,000$ images, while CD needs less time than De-Pois. For Fourclass and House Pricing datasets, De-Pois also takes the longest time. This is mainly due to the extra loading time of the mimic model of De-Pois. It is also noticed that the run-time overhead is largely affected by the dataset size where De-Pois spends longer time on CIFAR-10 dataset.

In addition, we report the training time of the synthetic data generation and the mimic model construction in De-Pois, so as to better understand the efficiency of De-Pois. For synthesizing every $500$ samples of CIFAR-10, Fourclass, and House Pricing datasets, De-Pois takes $544$ seconds, $27.1$ seconds, and $32.3$ seconds, respectively. It is noted that De-Pois takes longer time in training image dataset due to longer convergence time of the model. For mimic model construction on CIFAR-10, Fourclass, and House Pricing datasets, De-Pois takes $654$ seconds, $30.75$ seconds, and $179.5$ seconds for every $500$ samples, respectively.
We can also observe that the dimension of the training data has an impact on the training time.
CIFAR-10 and House Pricing datasets, which have higher dimensions, have much longer training time than that of Fourclass dataset.

\begin{table}[t]
	\centering
	\caption{Comparison of detection time (s).} \label{tab:detection_time}
	{\fontsize{8.5}{11.5}\selectfont
		\begin{tabular}{|c|c|c|c|l|}
			\hline
			Dataset                                         & \multicolumn{2}{c|}{CIFAR-10}          & \multicolumn{2}{c|}{Fourclass}     \\ \hline
			Methods                                         & De-Pois                       & CD        & De-Pois             & DUTI            \\ \hline
			Detection Time                                  & \multicolumn{1}{c|}{0.108} & 0.05      & 0.306            & 0.069           \\ \hline \hline
			Dataset                                         & \multicolumn{4}{c|}{House Pricing}                                          \\ \hline
			Methods                                         & De-Pois  & TRIM      & \multicolumn{2}{c|}{Sever}         \\ \hline
			\multicolumn{1}{|c|}{Detection Time}            & 0.068                      & 0.007     & \multicolumn{2}{c|}{0.027}         \\ \hline
			
		\end{tabular}
	}
\end{table}

\begin{table}[t]
	\centering
	\caption{Training time for De-Pois' components (s).} \label{tab:time overhead}
	{\fontsize{8.5}{11.5}\selectfont
		\begin{tabular}{|c|c|c|c|l|}
			\hline
			\diagbox{Component}{Training Time}{Dataset}                 & CIFAR-10                   & Fourclass & \multicolumn{2}{c|}{House Pricing} \\ \hline
			\multicolumn{1}{|l|}{Synthetic Data Generation} & 544                      & 27.1      & \multicolumn{2}{c|}{32.3}          \\ \hline
			\multicolumn{1}{|l|}{Mimic Model Construction}  & 654                      & 30.75      & \multicolumn{2}{c|}{179.5}         \\ \hline
		\end{tabular}
	}
\end{table}

\section{Conclusion} \label{conclusion}
In this paper, we have presented De-Pois, the first generic method for defending against data poisoning attacks. Given a small quantity of trusted clean data, we modify cGAN to obtain sufficient valid training data with a similar distribution of the trusted clean data.
We further utilize conditional WGAN-GP to train an effective mimic model, which can achieve comparable prediction performance with the target model.
By doing so, we can thus distinguish the poisoned data from the clean data by recognizing the prediction differences.
Experiment results on four realistic datasets against several typical kinds of poisoning attacks validate that De-Pois is attack-agnostic and very effective, and performs better in most cases compared with existing defense techniques.
We believe our work may deepen the understanding about the data poisoning attacking/defending mechanisms in practical settings and shed light on developing more efficient outlier detection techniques in broader areas.

%\balance
%\IEEEtriggeratref{27}
\bibliographystyle{IEEEtran}
\bibliography{mybib}

\begin{thebibliography}{10}
\providecommand{\url}[1]{#1}
\csname url@samestyle\endcsname
\providecommand{\newblock}{\relax}
\providecommand{\bibinfo}[2]{#2}
\providecommand{\BIBentrySTDinterwordspacing}{\spaceskip=0pt\relax}
\providecommand{\BIBentryALTinterwordstretchfactor}{4}
\providecommand{\BIBentryALTinterwordspacing}{\spaceskip=\fontdimen2\font plus
\BIBentryALTinterwordstretchfactor\fontdimen3\font minus
  \fontdimen4\font\relax}
\providecommand{\BIBforeignlanguage}[2]{{%
\expandafter\ifx\csname l@#1\endcsname\relax
\typeout{** WARNING: IEEEtran.bst: No hyphenation pattern has been}%
\typeout{** loaded for the language `#1'. Using the pattern for}%
\typeout{** the default language instead.}%
\else
\language=\csname l@#1\endcsname
\fi
#2}}
\providecommand{\BIBdecl}{\relax}
\BIBdecl

\bibitem{RN208}
Y.~LeCun, Y.~Bengio, and G.~Hinton, ``Deep learning,'' \emph{Nature}, vol. 521,
  pp. 436--444, 2015.

\bibitem{7873244}
M.~{De Cock}, R.~{Dowsley}, C.~{Horst}, R.~{Katti}, A.~C.~A. {Nascimento},
  W.~{Poon}, and S.~{Truex}, ``Efficient and private scoring of decision trees,
  support vector machines and logistic regression models based on
  pre-computation,'' \emph{IEEE Transactions on Dependable and Secure
  Computing}, vol.~16, no.~2, pp. 217--230, 2019.

\bibitem{tramer:2016stealing}
F.~Tram{\`e}r, F.~Zhang, A.~Juels, M.~K. Reiter, and T.~Ristenpart, ``Stealing
  machine learning models via prediction {APIs},'' in \emph{Proceedings of
  USENIX Security}, 2016, pp. 601--618.

\bibitem{shokri:2017membership}
R.~Shokri, M.~Stronati, C.~Song, and V.~Shmatikov, ``Membership inference
  attacks against machine learning models,'' in \emph{Proceedings of IEEE
  S\&P}, 2017, pp. 3--18.

\bibitem{fredrikson:2015model}
M.~Fredrikson, S.~Jha, and T.~Ristenpart, ``Model inversion attacks that
  exploit confidence information and basic countermeasures,'' in
  \emph{Proceedings of ACM CCS}, 2015, pp. 1322--1333.

\bibitem{hitaj:2017deep}
B.~Hitaj, G.~Ateniese, and F.~Perez-Cruz, ``Deep models under the {GAN}:
  information leakage from collaborative deep learning,'' in \emph{Proceedings
  of ACM CCS}, 2017, pp. 603--618.

\bibitem{biggio:2013evasion}
B.~Biggio, I.~Corona, D.~Maiorca, B.~Nelson, N.~{\v{S}}rndi{\'c}, P.~Laskov,
  G.~Giacinto, and F.~Roli, ``Evasion attacks against machine learning at test
  time,'' in \emph{Proceedings of ECML PKDD}, 2013, pp. 387--402.

\bibitem{8949445}
J.~{Wu}, B.~{Chen}, W.~{Luo}, and Y.~{Fang}, ``Audio steganography based on
  iterative adversarial attacks against convolutional neural networks,''
  \emph{IEEE Transactions on Information Forensics and Security}, vol.~15, pp.
  2282--2294, 2020.

\bibitem{Battista-et-al:poisoning}
B.~Battista, N.~Blaine, and L.~Pavel, ``Poisoning attacks against support
  vector machines,'' in \emph{Proceedings of ICML}, 2012, pp. 1467--1474.

\bibitem{alfeld:2016data}
S.~Alfeld, X.~Zhu, and P.~Barford, ``Data poisoning attacks against
  autoregressive models,'' in \emph{Proceedings of AAAI}, 2016, pp. 1452--1458.

\bibitem{mei:2015using}
S.~Mei and X.~Zhu, ``Using machine teaching to identify optimal training-set
  attacks on machine learners.'' in \emph{Proceedings of AAAI}, 2015, pp.
  2871--2877.

\bibitem{munoz:2017towards}
L.~Mu{\~n}oz-Gonz{\'a}lez, B.~Biggio, A.~Demontis, A.~Paudice, V.~Wongrassamee,
  E.~C. Lupu, and F.~Roli, ``Towards poisoning of deep learning algorithms with
  back-gradient optimization,'' in \emph{Proceedings of ACM Workshop on AISec},
  2017, pp. 27--38.

\bibitem{koh:2018stronger}
P.~W. Koh, J.~Steinhardt, and P.~Liang, ``Stronger data poisoning attacks break
  data sanitization defenses,'' \emph{CoRR, arXiv: 1811.00741}, 2018.

\bibitem{gu:2017badnets}
T.~Gu, B.~Dolan-Gavitt, and S.~Garg, ``Badnets: Identifying vulnerabilities in
  the machine learning model supply chain,'' \emph{CoRR, arXiv: 1708.06733},
  2017.

\bibitem{chen-et-al:2017targeted}
X.~Chen, C.~Liu, B.~Li, K.~Lu, and D.~Song, ``Targeted backdoor attacks on deep
  learning systems using data poisoning,'' \emph{CoRR, arXiv: 1712.05526},
  2017.

\bibitem{8854834}
C.~{Chen}, X.~{Zhao}, and M.~C. {Stamm}, ``Generative adversarial attacks
  against deep-learning-based camera model identification,'' \emph{IEEE
  Transactions on Information Forensics and Security}, to appear. DOI:
  10.1109/TIFS.2019.2945198.

\bibitem{paudice:2018detection}
A.~Paudice, L.~Mu{\~n}oz-Gonz{\'a}lez, A.~Gyorgy, and E.~C. Lupu, ``Detection
  of adversarial training examples in poisoning attacks through anomaly
  detection,'' \emph{CoRR, arXiv: 1802.03041}, 2018.

\bibitem{liu:2018fine}
K.~Liu, B.~Dolan-Gavitt, and S.~Garg, ``Fine-pruning: Defending against
  backdooring attacks on deep neural networks,'' in \emph{Proceedings of RAID},
  2018, pp. 273--294.

\bibitem{carnerero:2020regularisation}
J.~Carnerero-Cano, L.~Mu{\~n}oz-Gonz{\'a}lez, P.~Spencer, and E.~C. Lupu,
  ``Regularisation can mitigate poisoning attacks: A novel analysis based on
  multiobjective bilevel optimisation,'' \emph{CoRR, arXiv: 2003.00040}, 2020.

\bibitem{peri:2019deep}
N.~Peri, N.~Gupta, W.~Ronny~Huang, L.~Fowl, C.~Zhu, S.~Feizi, T.~Goldstein, and
  J.~P. Dickerson, ``Deep {$k$-NN} defense against clean-label data poisoning
  attacks,'' \emph{CoRR, arXiv: 1909.13374}, 2019.

\bibitem{Yang:2017GenerativePA}
C.~Yang, Q.~Wu, H.~Li, and Y.~Chen, ``Generative poisoning attack method
  against neural networks,'' \emph{CoRR, arXiv: 1703.01340}, 2017.

\bibitem{Jagielski:2018ManipulatingML}
M.~Jagielski, A.~Oprea, B.~Biggio, C.~Liu, C.~Nita-Rotaru, and B.~Li,
  ``Manipulating machine learning: Poisoning attacks and countermeasures for
  regression learning,'' in \emph{Proceedings of IEEE S\&P}, 2018, pp. 19--35.

\bibitem{Biggio:2011SupportVM}
B.~Biggio, B.~Nelson, and P.~Laskov, ``Support vector machines under
  adversarial label noise,'' in \emph{Proceedings of ACML}, 2011, pp. 97--112.

\bibitem{miao:2018attack}
C.~Miao, Q.~Li, L.~Su, M.~Huai, W.~Jiang, and J.~Gao, ``Attack under disguise:
  An intelligent data poisoning attack mechanism in crowdsourcing,'' in
  \emph{Proceedings of WWW}, 2018, pp. 13--22.

\bibitem{Goodfellow:2014}
I.~J. Goodfellow, J.~Pouget-Abadie, M.~Mirza, B.~Xu, D.~Warde-Farley, S.~Ozair,
  A.~C. Courville, and Y.~Bengio, ``Generative adversarial networks,'' in
  \emph{Proceedings of NeurIPS}, 2014, pp. 2672--2680.

\bibitem{Mirza-Osindero:2014CGAN}
M.~Mirza and S.~Osindero, ``Conditional generative adversarial nets,''
  \emph{CoRR, arXiv: 1411.1784}, 2014.

\bibitem{gulrajani:2017improved}
I.~Gulrajani, F.~Ahmed, M.~Arjovsky, V.~Dumoulin, and A.~C. Courville,
  ``Improved training of {Wasserstein GANs},'' in \emph{Proceedings of
  NeurIPS}, 2017, pp. 5767--5777.

\bibitem{Steinhardt:2017CertifiedDF}
J.~Steinhardt, P.~W. Koh, and P.~Liang, ``Certified defenses for data poisoning
  attacks,'' in \emph{Proceedings of NeurIPS}, 2017, pp. 3520--3532.

\bibitem{zhang:2018training}
X.~Zhang, X.~Zhu, and S.~J. Wright, ``Training set debugging using trusted
  items,'' in \emph{Proceedings of AAAI}, 2018.

\bibitem{diakonikolas:2019sever}
I.~Diakonikolas, G.~Kamath, D.~Kane, J.~Li, J.~Steinhardt, and A.~Stewart,
  ``Sever: A robust meta-algorithm for stochastic optimization,'' in
  \emph{Proceedings of ICML}, 2019, pp. 1596--1606.

\bibitem{zhu:2019transferable}
C.~Zhu, W.~R. Huang, A.~Shafahi, H.~Li, G.~Taylor, C.~Studer, and T.~Goldstein,
  ``Transferable clean-label poisoning attacks on deep neural nets,'' in
  \emph{Proceedings of ICML}, 2019, pp. 7614--7623.

\bibitem{munoz:2019poisoning}
L.~Mu{\~n}oz-Gonz{\'a}lez, B.~Pfitzner, M.~Russo, J.~Carnerero-Cano, and E.~C.
  Lupu, ``Poisoning attacks with generative adversarial nets,'' \emph{CoRR,
  arXiv: 1906.07773}, 2019.

\bibitem{shafahi:2018poison}
A.~Shafahi, W.~R. Huang, M.~Najibi, O.~Suciu, C.~Studer, T.~Dumitras, and
  T.~Goldstein, ``Poison frogs! targeted clean-label poisoning attacks on
  neural networks,'' in \emph{Proceedings of NeurIPS}, 2018, pp. 6103--6113.

\bibitem{Paudice:2018LabelSA}
A.~Paudice, L.~Mu{\~n}oz-Gonz{\'a}lez, and E.~C. Lupu, ``Label sanitization
  against label flipping poisoning attacks,'' in \emph{Proceedings of ECML
  PKDD}, 2018, pp. 5--15.

\bibitem{biggio:2018wild}
B.~Biggio and F.~Roli, ``Wild patterns: Ten years after the rise of adversarial
  machine learning,'' \emph{Pattern Recognition}, vol.~84, pp. 317--331, 2018.

\bibitem{tran:2017bayesian}
T.~Tran, T.~Pham, G.~Carneiro, L.~Palmer, and I.~Reid, ``A bayesian data
  augmentation approach for learning deep models,'' in \emph{Proceedings of
  NeurIPS}, 2017, pp. 2797--2806.

\bibitem{Tanner:2012tool}
M.~A. Tanner, \emph{Tools for statistical inference: observed data and data
  augmentation methods}.\hskip 1em plus 0.5em minus 0.4em\relax Springer, 2012.

\bibitem{arjovsky:2017wasserstein}
M.~Arjovsky, S.~Chintala, and L.~Bottou, ``Wasserstein generative adversarial
  networks,'' in \emph{Proceedings of ICML}, 2017, pp. 214--223.

\bibitem{gao:2019strip}
Y.~Gao, C.~Xu, D.~Wang, S.~Chen, D.~C. Ranasinghe, and S.~Nepal, ``{STRIP}: A
  defence against trojan attacks on deep neural networks,'' in
  \emph{Proceedings of ACSAC}, 2019, pp. 113--125.

\bibitem{Udeshi:2019ModelAD}
S.~Udeshi, S.~Peng, G.~Woo, L.~Loh, L.~Rawshan, and S.~Chattopadhyay, ``Model
  agnostic defence against backdoor attacks in machine learning,'' \emph{CoRR,
  arXiv: 1908.02203}, 2019.

\bibitem{salimans:2016improved}
T.~Salimans, I.~Goodfellow, W.~Zaremba, V.~Cheung, A.~Radford, and X.~Chen,
  ``Improved techniques for training {GAN}s,'' in \emph{Proceedings of
  NeurIPS}, 2016, pp. 2234--2242.

\bibitem{heusel:2017gans}
M.~Heusel, H.~Ramsauer, T.~Unterthiner, B.~Nessler, and S.~Hochreiter, ``{GAN}s
  trained by a two time-scale update rule converge to a local {Nash}
  equilibrium,'' in \emph{Proceedings of NeurIPS}, 2017, pp. 6626--6637.

\bibitem{dowson:1982frechet}
D.~Dowson and B.~Landau, ``The fr{\'e}chet distance between multivariate normal
  distributions,'' \emph{Journal of Multivariate Analysis}, vol.~12, no.~3, pp.
  450--455, 1982.

\end{thebibliography}

\begin{IEEEbiography}[{\includegraphics[width=1in,height=1.25in,clip,keepaspectratio]{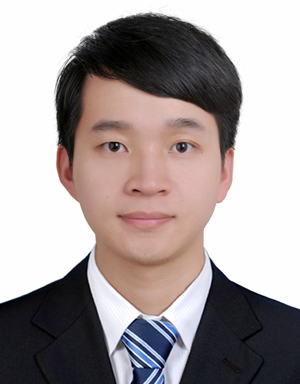}}]{Jian Chen} (S'20)
received B.S. degree from Hubei University of Technology in 2014 and the M.S. degree from Huazhong University of Science and Technology in 2018. He is currently working toward the Ph.D. degree in School of Electronic Information and Communications, Huazhong University of Science and Technology, China. His recent research interests focus on machine learning and data privacy. He is a student member of IEEE.
\end{IEEEbiography}

\begin{IEEEbiography}[{\includegraphics[width=1in,height=1.25in,clip,keepaspectratio]{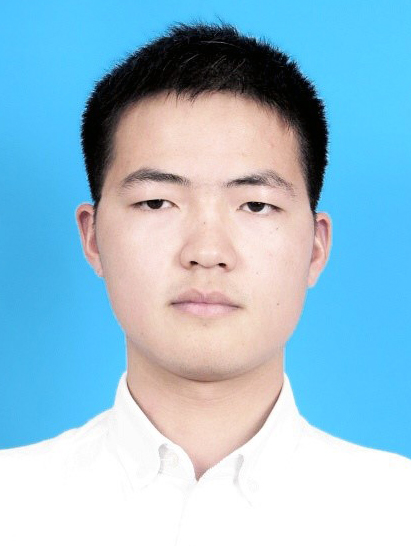}}]{Xuxin Zhang}
received the B.E. degree from University of Electronic Science and Technology of China, China, in 2019. He is currently pursuing the M.S. degree in Electronics and Information Engineering at Huazhong University of Science and Technology, China. His research interests include data poisoning attacks and recommender systems.
\end{IEEEbiography}

%\vspace{-1cm}
\begin{IEEEbiography}[{\includegraphics[width=1in,height=1.25in,clip,keepaspectratio]{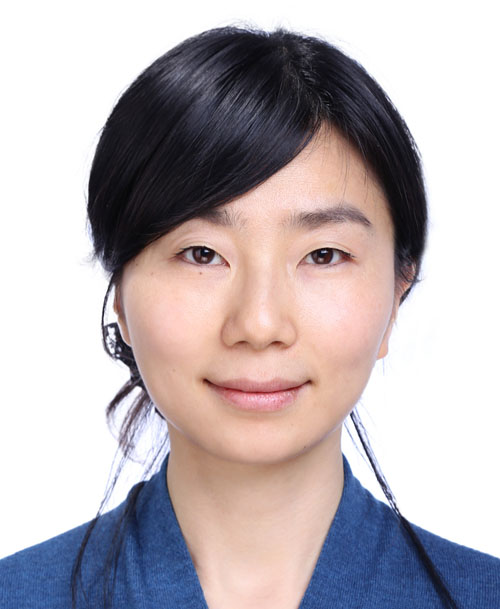}}]{Rui Zhang} (M'12)
is an Associate Professor in School of Computer Science and Technology at Wuhan University of Technology, China. She received the M.S. degree and Ph.D. degree in Computer Science from Huazhong University of Science and Technology, China. From 2013 to 2014, she was a Visiting Scholar with the College of Computing, Georgia Institute of Technology, USA. Her research interests include machine learning, mobile computing and data analytics.
\end{IEEEbiography}

%\vspace{-1cm}
\begin{IEEEbiography}[{\includegraphics[width=1in,height=1.25in,clip,keepaspectratio]{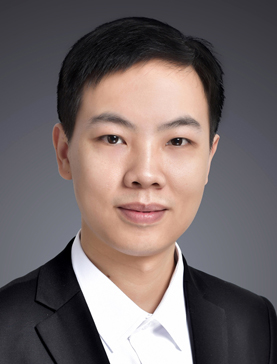}}]{Chen Wang}
(S'10-M'13-SM'19) received the B.S. and Ph.D. degrees from the Department of Automation, Wuhan University, China, in 2008 and 2013, respectively. From 2013 to 2017, he was a postdoctoral research fellow in the Networked and Communication Systems Research Lab, Huazhong University of Science and Technology, China. Thereafter, he joined the faculty of Huazhong University of Science and Technology where he is currently an associate professor. His research interests are in the broad areas of wireless networking, Internet of Things, and mobile computing, with a recent focus on privacy issues in wireless and mobile systems. He is a senior member of IEEE and ACM.
\end{IEEEbiography}

\vspace{-0.8cm}
\begin{IEEEbiography}[{\includegraphics[width=1in,height=1.25in,clip,keepaspectratio]{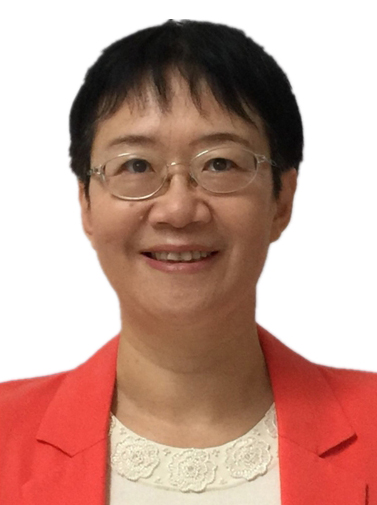}}]{Ling Liu}
(F'15) is currently a Professor at the School of Computer Science, Georgia Institute of Technology, Atlanta, GA, USA. She directs the research programs at the Distributed Data Intensive Systems Lab (DiSL), examining various aspects of large-scale big data systems and analytics, including performance, availability, security, privacy, and trust. Her current research is sponsored primarily by the National Science
Foundation and IBM. She has published over 300 international journal and conference articles. Dr. Liu was a recipient of the IEEE Computer Society Technical Achievement Award in 2012 and the Best Paper Award from numerous top venues, including ICDCS, WWW, IEEE Cloud, IEEE ICWS, and ACM/IEEE CCGrid. She served as the general chair and the PC chair for numerous IEEE and ACM conferences in big data, distributed computing, cloud computing, data engineering, very large databases, and the World Wide Web fields. She served as the Editor-in-Chief for the IEEE Transactions on Services Computing from 2013 to 2016. She is the Editor-in-Chief of the ACM Transactions on Internet Technology.
\end{IEEEbiography}

\end{document}